\documentclass[journal]{IEEEtran} 
\usepackage{setspace}
\usepackage{url}  
\usepackage{mleftright}
\usepackage{cite}   
\usepackage[pdftex]{graphicx} 
\graphicspath{ {./image/} } 
\usepackage[cmex10]{amsmath}
\usepackage{array} 
\ifCLASSOPTIONcompsoc
\usepackage[caption=false, font=normalsize, labelfont=sf, textfont=sf]{subfig}
\else
\usepackage[caption=false, font=footnotesize]{subfig}
\usepackage{subfig}
\usepackage{diagbox}
\usepackage{booktabs}
\usepackage{ wasysym }
\usepackage[T1]{fontenc} 
\allowdisplaybreaks
\usepackage{color}
\usepackage{float}
\usepackage{fixmath}
\usepackage{stfloats}
\usepackage{arydshln} 
\usepackage[]{changes}
\usepackage{textcomp} 
\usepackage{adjustbox} 
\usepackage{textcase}
\usepackage{graphicx}
\usepackage{enumerate}
\usepackage{footnote}
\usepackage{xparse}
\usepackage{amssymb}
\usepackage{amsmath}
\usepackage{amsthm}
\usepackage{amsfonts}
\RequirePackage{xspace}
\RequirePackage{graphics}
\RequirePackage{textcomp}
\usepackage{keyval}
\usepackage{xspace}
\usepackage{mathrsfs}
\usepackage{paralist}
\usepackage{listings}
\usepackage{multirow}
\usepackage{array}
\usepackage{stmaryrd}
\usepackage{textcase}
\usepackage{makecell}
\usepackage{stackengine}%
\usepackage[utf8]{inputenc}
\usepackage[english]{babel}

\usepackage{flushend}
\usepackage{balance}
\usepackage{etoolbox}

\usepackage{bm}
\usepackage{amsmath}
\usepackage{xcolor,cite,etoolbox}

\newtheorem{definition}{Definition}

\newcommand{\seclabel}[1]{\label{sec:#1}}

\renewcommand{\eqref}[1]{~(\ref{eq:#1})}

\newcommand{\abf}{\boldsymbol{a}}
\newcommand{\Abf}{\boldsymbol{A}}

\newcommand{\Cbf}{\boldsymbol{C}}

\newcommand{\Ccal}{\mathcal{C}}
\newcommand{\Cbb}{\mathbb{C}}

\newcommand{\Fcal}{\mathcal{F}}

\newcommand{\Gbf}{\boldsymbol{G}}
\newcommand{\Gcal}{\mathcal{G}}

\newcommand{\Hcal}{\mathcal{H}}
\newcommand{\Hbb}{\mathbb{H}}

\newcommand{\Lbf}{\boldsymbol{L}}
\newcommand{\Lcal}{\mathcal{L}}

\newcommand{\Ncal}{\mathcal{N}}

\newcommand{\pbf}{\boldsymbol{p}}

\newcommand{\qbf}{\boldsymbol{q}}

\newcommand{\Rbb}{\mathbb{R}}

\newcommand{\sbf}{\boldsymbol{s}}

\newcommand{\Sbb}{\mathbb{S}}

\newcommand{\vbf}{\boldsymbol{v}}

\newcommand{\Ybf}{\boldsymbol{Y}}

\newcommand{\diagrm}{\mathrm{diag}}

\makeatletter
\DeclareRobustCommand{\cev}[1]{%
  \mathpalette\do@cev{#1}%
}
\newcommand{\do@cev}[2]{%
  \fix@cev{#1}{+}%
  \reflectbox{$\m@th#1\vec{\reflectbox{$\fix@cev{#1}{-}\m@th#1#2\fix@cev{#1}{+}$}}$}%
  \fix@cev{#1}{-}%
}
\newcommand{\fix@cev}[2]{%
  \ifx#1\displaystyle
    \mkern#2 1mu
  \else
    \ifx#1\textstyle
      \mkern#2 3mu
    \else
      \ifx#1\scriptstyle
        \mkern#2 2mu
      \else
        \mkern#2 2mu
      \fi
    \fi
  \fi
}

\usepackage{stackengine}
\stackMath

\makeatletter 
\pretocmd\@bibitem{\color{black}\csname keycolor#1\endcsname}{}{\fail}
\newcommand\citecolor[1]{\@namedef{keycolor#1}{\color{blue}}}
\makeatother

\begin{document}
	\title{Enhanced Modeling of Contingency Response in Security-constrained Optimal Power Flow}
	
\author{Tuncay~Altun, 
Ramtin~Madani, 
Alper Atamt\"{u}rk, 
Ross Baldick,
and~Ali~Davoudi 
		
\thanks{This work was supported by the National Science Foundation and Department of Energy under awards ECCS-1809454 and DE-AR0001086, respectively. Tuncay Altun, Ramtin Madani, and Ali Davoudi are with the University of Texas at Arlington. Alper Atamt\"{u}rk is with University of California, Berkeley. Ross Baldick is with the University of Texas at Austin (e-mail: tuncay.altun@uta.edu; ramtin.madani@uta.edu; atamturk@berkeley.edu, baldick@mail.utexas.edu, davoudi@uta.edu). }}
	\maketitle
	%
	\begin{abstract}
		This paper provides an enhanced modeling of the contingency response that collectively reflects high-fidelity physical and operational characteristics of power grids. 
Integrating active and reactive power contingency responses into the security-constrained optimal power flow (SCOPF) problem is challenging, due to the nonsmoothness and nonconvexity of feasible sets in consequence of piece-wise curves representing generator characteristics. We introduce a continuously-differentiable model using functions that closely resemble PV/PQ switching and the generator contingency response. These models enforce physical and operational limits by optimally allocating active power imbalances among available generators and deciding the bus type to switch from the PV type to the PQ type. The efficacy of this method is numerically validated on the IEEE 30-bus, 300-bus, and 118-bus systems with 12, 10, and 100 contingencies, respectively.

%
%

	\end{abstract}
	
	\IEEEpeerreviewmaketitle
	
	\newcommand{\mysmallarraydecl}{\renewcommand{
			\IEEEeqnarraymathstyle}{\scriptscriptstyle}%
		\renewcommand{\IEEEeqnarraytextstyle}{\scriptsize}
		\settowidth{\normalbaselineskip}{\scriptsize
			\hspace{\baselinestretch\baselineskip}}%
		\setlength{\baselineskip}{\normalbaselineskip}%
		\setlength{\jot}{0.25\normalbaselineskip}%
		\setlength{\arraycolsep}{2pt}}

\section{Introduction}
\seclabel{intro}
%

\IEEEPARstart{P}{ower} flow analysis underpins many static and dynamic applications, including stability analysis, optimal power flow, contingency analysis, etc. Accurate power flow solutions ensure the generation-demand balance under all circumstances. However, since transmission losses cannot be identified a priori, the total power needed to supply a known demand remains unpredictable \cite{investigate_dsb}. The common practice is to assume that there exist at least one slack bus, where active power generation can be readjusted to compensate for imbalances\cite{book1, l3}. Power flow models that are based on multiple slack buses alleviate the burden of a pre-specified single slack bus by dispatching the imbalance among participating sources  \cite{incre_ping}. This approach better mimics the operation of power systems provided that participation factors of generators are accurately determined. These participation factors can be appointed based on machine inertia \cite{unbund_ilic}, governor droop characteristics \cite{droop1,virtual}, frequency control \cite{slack_sel}, and economic dispatch \cite{ecd_part,incre_meisel, jang2005modified}. Another major aspect of post-contingency analysis is the determination of reactive power dispatch and voltage magnitudes \cite{novelopf}. When the reactive power limit of a generator is reached, it cannot maintain predefined voltage settings, and the bus type should inevitably switch from PV bus to PQ bus \cite{l11}. The active and reactive power limits are handled via controller design \cite{l8}, \cite{upf} by sacrificing the optimal operation. Moreover, possible component outage, i.e. generator or line, are rarely studied in the literature \cite{l7}.

 

Security-constrained optimal power flow (SCOPF) formulations focus on the optimization of a robust power dispatch with respect to the outage of arbitrary sets of generators or lines \cite{l2,l3}. 
Corrective models of SCOPF allow limited adjustment of operating points in response to contingencies, i.e., redispatch, in post-contingency scenarios \cite{l5}. 
In this case, to account for a realistic post-contingency behavior of generators, additional constraints have to be incorporated into corrective models of SCOPF formulation \cite{l10}. These additional constraints are based on nonsmooth, i.e., discontinuous, curves of real and reactive power responses to avoid prevent both active and reactive power violations \cite{l9}. However, nonsmooth models prevent the use of power flow algorithms due to non-differentiability  \cite{l12}. This problem, and its possible adverse consequences, such as an increased iteration count and convergence to an unstable region, have been discussed in \cite{l9}.

Alternative differentiable power flow models using hyperbolic and sigmoid functions \cite{l10}, patching functions with complementary homotopy methods \cite{l12}, and discrete and continuous auxiliary variables, are discussed in \cite{l11}. 
In this paper, we introduce several continuously-differentiable models that respect high-fidelity physical models of generators considering an extensive list of contingency scenarios. These models prevent physical and operational violations by means of optimally allocating active power imbalances among available generators and deciding the bus type, i.e., PV/PQ switching, respectively. While the employed local search solver fail to converge using continuous complementarity condition from \cite{l11}, the proposed model enable the recovery of fully feasible solutions in all of the simulated cases. 

The rest of this paper has the following organization. Section II discusses the preliminary materials. Section III elaborates the enhanced modeling of the generator response including representation of a distributed slack bus as well as active and reactive power contingency response models for a generator. Section IV introduces several continuously-differentiable models that account for active and reactive power contingency responses in the SCOPF formulation. In Section V, the proposed enhanced modeling of generator response in SCOPF solution is verified through a numerical benchmark system. Section VI concludes the paper.

\section{Notations and Power Grid Terminologies}
\subsection{Notations} 

Bold lowercase and uppercase letters (e.g., $\abf,~ \Abf$) represent vectors and matrices, respectively. $\bm{1}$ and $\bm{0}$ refer to vectors with all elements as 1 and 0, respectively. The sets of complex and real numbers are shown with $\Cbb$ and $\Rbb$, respectively. $\Hbb^n$ and $\Sbb^n$ represent the hermitian and symmetric matrices of size $n \times n$, respectively. $\mathrm{imag}\{\cdot\}$ and $\mathrm{real}\{\cdot\}$ define the imaginary and real parts of a complex matrix or number, respectively. Superscripts $\!(\cdot)^{\!\top}\!$ and $\!(\cdot)^{\!\ast}\!$ stand for the transpose and conjugate transpose operator, respectively.  $|\cdot|$ represents the cardinality of a set or the absolute/magnitude value of a vector/scalar. 
$\diagrm\{\cdot\}$ forms a vector  using diagonal entries of a given matrix. $[\cdot]$ composes a diagonal matrix from a given vector. 
  \begin{figure} 
 	\centering  
 	\includegraphics[width=\columnwidth]{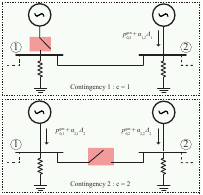}
 	\caption{Contingency scenario examples for a portion of the power grid. The base case refers to the scenario 0, i.e., no contingency. The red labels highlight the failure of a generator and a transmission line. In Scenario 1, generator 1 fails while generator 2 readjusts its output according to weight $\alpha_{1,2}$. Scenario 2 illustrates a transmission line failure, while generators 1 and 2 contribute to the power re-dispatch $\Delta_2$.}
 	\label{fig:1}
 \end{figure}
 \subsection{Power Grid Terminologies} 
 
Figure~\ref{fig:1} shows a snapshot of a power grid, where any single bus can accommodate an arbitrary number of generators and loads. The set of contingency scenarios is referred to as $\Ccal$, with each positive member accounting for the outage of, at least, one network component, e.g., a generator or a transmission line. Herein, the \textit{base case scenario}, i.e., normal operation with no contingency, is shown by $0 \in \Ccal$. For the rest of the scenarios, $c \in \Ccal$, grid terminologies are detailed as:
\begin{itemize}
\item {\bf Buses and Lines:} The transmission grid is structured using a directed graph $\Hcal=\left(\Ncal,\Lcal\right)$, where  the sets of buses and lines are denoted by $\Ncal$ and $\Lcal$, respectively. Let   $\vec{\Lbf}_c,\cev{\Lbf}_c\in\{0,1\}^{|\Lcal|\times|\Ncal|}$  be the pairs for the {\it from} and {\it to} line-incidence matrices in a contingency case, respectively. $\vec{L}_{c,_{li}} = 1$ and $\cev{L}_{c,_{li}}=1$ for every $l\in\mathcal{L}$, iff the transmission line $l$ starts at bus $i$, and vice versa, respectively. Matrices $\Ybf_c \in \Cbb^{|\Ncal| \times |\Ncal|}$, $\vec{\Ybf}_c,\cev{\Ybf}_c \in \Cbb^{|\Lcal| \times |\Ncal|}$ respectively denote the bus-admittance, and the {\it from} and {\it to} line-admittance matrices (See \cite{zimmerman2016matpower} for the definition of admittance matrices). Let $\sbf^{\rm{max}}_c\in(\Cbb\cup\{\infty\})^{|\Lcal|}$ be the vector of apparent power flow limits on transmission lines. Define $\vbf_0$, $\vbf_c \in \Cbb^{|\Ncal|}$ as the vectors of complex nodal voltages in the base and contingency cases, respectively. Define $\vbf^{\rm{max}},\vbf^{\rm{min}}\in\Cbb^{|\Ncal|}$ as the vectors of the maximum and minimum voltage magnitudes, respectively.
\item {\bf Generators/Loads:} 
Let $\Gcal$ be the set of generators and $\Gbf_c \in \{0,1\}^{|\Gcal|\times|\Ncal|}$ as the generator incidence matrix in a contingency case; $(g,i)$ element is $1$ iff generator $g\in \Gcal$ is located at the bus $i\in \Ncal$ and not outed in the event of contingency $c$.   $\Cbf_c \in \{0,1\}^{|\Gcal|\times|\Gcal|}$ denotes a diagonal incidence matrix whose $(g,g)$ element is $1$, iff the generator $g\in\Gcal$ is in service in the event of contingency $c$.  $\sbf^{\rm{dem}}_c\in\Cbb$ represents the vectors of apparent power demand. Let $\sbf^{\rm{gen}}_0\in\Cbb^{|\Gcal|}$, and $\pbf^{\rm{gen}}_0,\qbf^{\rm{gen}}_0\in\Rbb^{|\Gcal|}$, respectively, represent the vectors of apparent, active, and reactive power generations in the base case, while for every $c\in\Ccal$ $\sbf^{\rm{gen}}_c\in\Cbb^{|\Gcal|}$, and $\pbf^{\rm{gen}}_c,\qbf^{\rm{gen}}_c\in\Rbb^{|\Gcal|}$ represent the corresponding post-contingency power generation vectors. Define $\pbf^{\rm{max}}_c,\qbf^{\rm{max}}_c\in\Rbb^{|\Gcal|}$
and $\pbf^{\rm{min}}_c,\qbf^{\rm{min}}_c\in\Rbb^{|\Gcal|}$ as the vectors of the maximum and minimum active and reactive power generations, respectively.

\end{itemize}	
	\section{Enhanced Modeling of a Generator Response}
\begin{figure} 
	\centering
	\subfloat(a){%
		\includegraphics[width=0.40\linewidth]{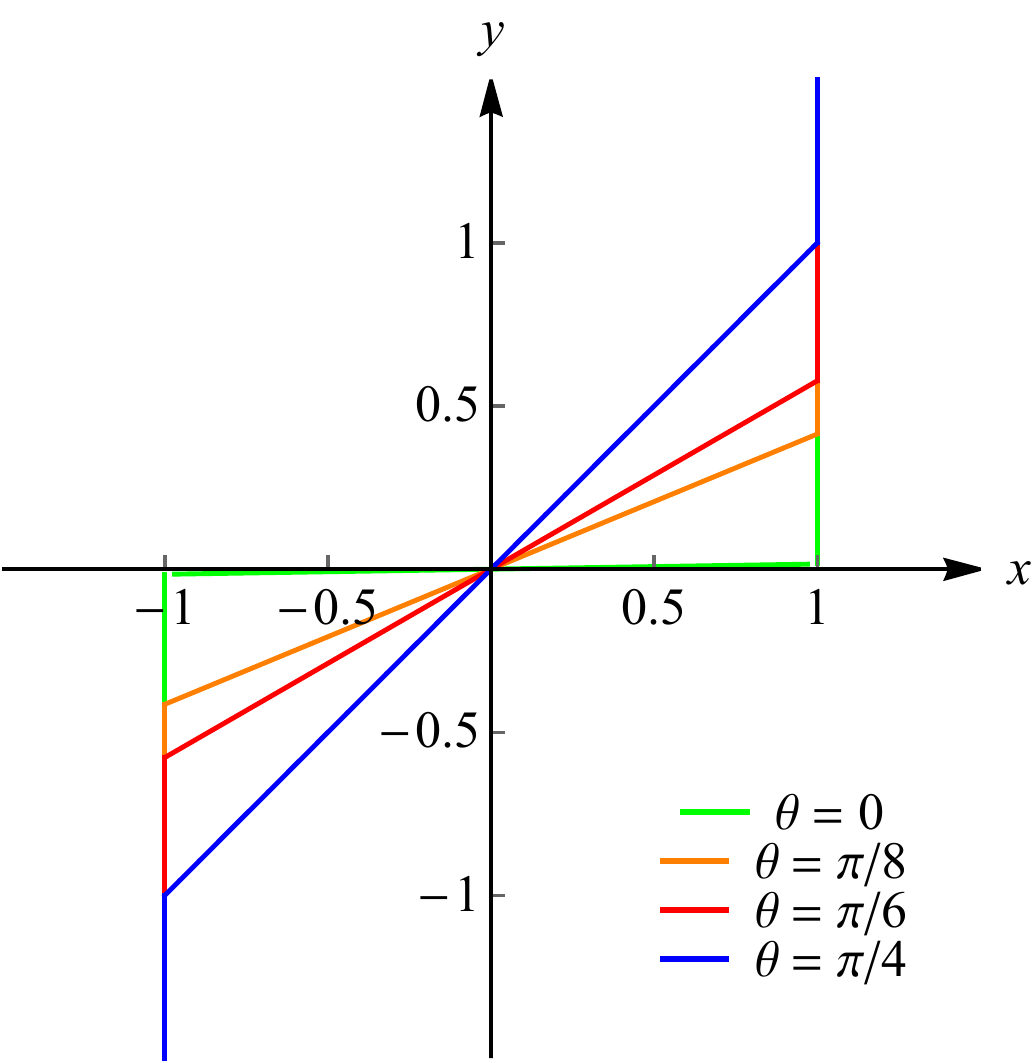}}
	\hfill
	\subfloat(b){%
		\includegraphics[width=0.40\linewidth]{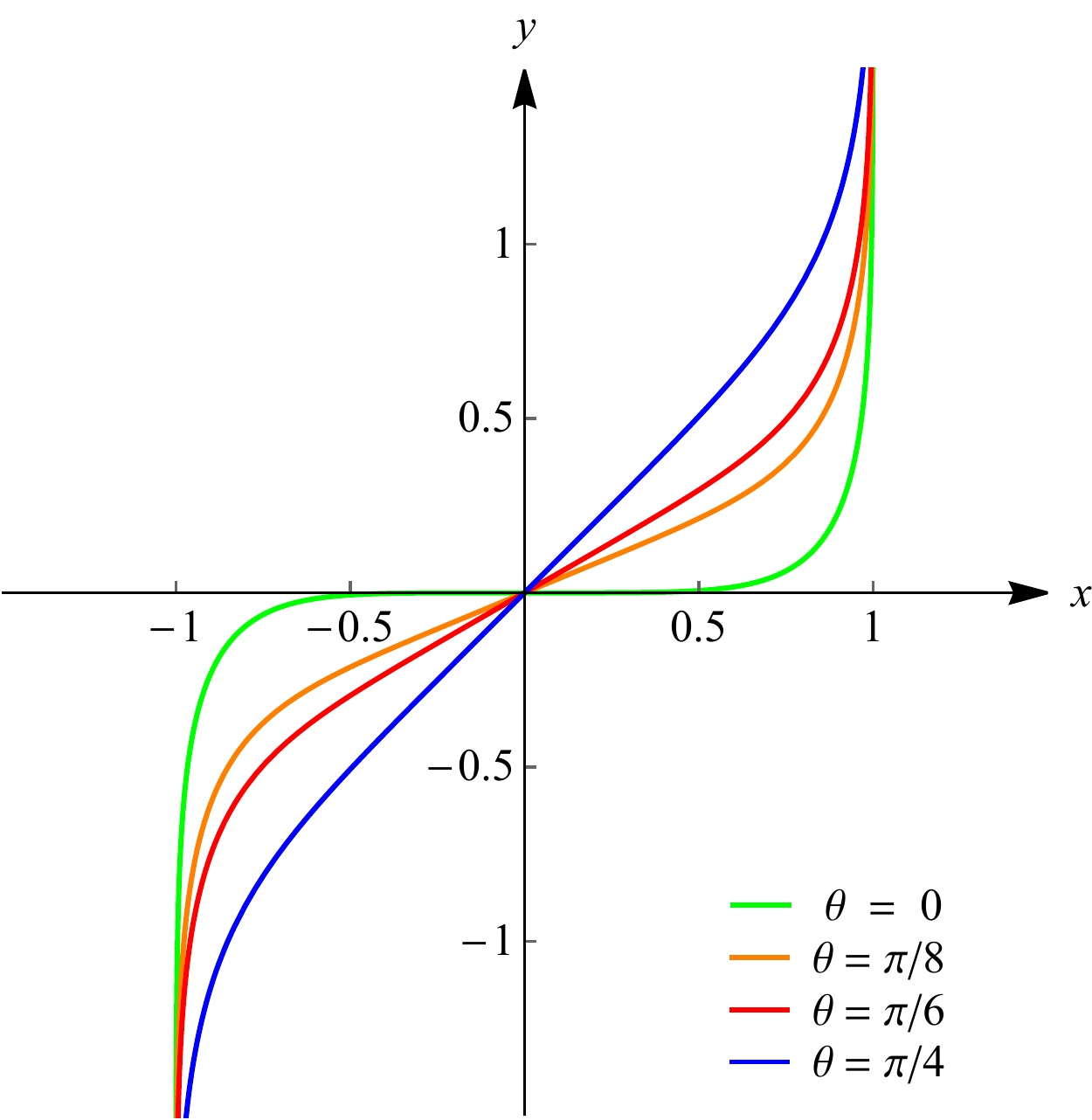}}
	\caption{The characteristics of (a) $(x,y) \in \Fcal_{\theta}$ and (b) $(x,y) \in \Fcal^{\mathrm{smooth}}_{\theta}$ for different slope values, $\theta=0$, $\theta=\pi/8$, $\theta=\pi/6$, $\theta=\pi/4$.}
	\label{fig2} 
\end{figure}

To capture the nonlinear characteristics of a power system, such as PV/PQ switching and generator contingency response, and to streamline these policies, we define the following set.
\begin{definition}
For $\theta \in [0, \pi/4)$, define $\Fcal_{\theta}\subseteq \mathbb{R}^2$ as
\begin{align} 
&\Fcal_{\theta}  \triangleq\Big\{\;(x,y)\in\mathbb{R}^2\;|\;-1 \leq x \leq 1\nonumber\\
&\;\wedge\;\min\{\max\left\{0,y-\mathrm{tan}(\theta)x\},\max\{0,1-x\}\right\}=0\nonumber\\
&\;\wedge\;\min\{\max\left\{0,\mathrm{tan}(\theta)x-y\},\max\{0,1+x\}\right\}=0\;\Big\}, \label{eq:set_orig}
\end{align}
\end{definition}
\noindent where $\theta$ denotes the slope for the segment
within the interval $[-1,1]$. Herein, each pair ($x$, $y$) represents the coordinates on a piecewise-smooth curve. In the following section, we will provide a smooth version of $\Fcal_{\theta}$ in order to facilitate local search. This is illustrated in Figure \ref{fig2}.
\subsection{Generator Active Power Contingency Response}
For now, assume that there is no limit imposed on post-contingency active power injections, i.e.,
\begin{align}
p^{\max}_{c,g} = -p^{\min}_{c,g} = \infty \qquad \forall c \in \Ccal \setminus{0}, g\in\Gcal. \label{111}
\end{align}
Given assumption (\ref{111}), the active power imbalance due to changes in network configuration, transmission losses, or load profile is distributed among operational generators as
\begin{equation}
p^{\mathrm{gen}}_{c,g}=p^{\mathrm{gen}}_{0,g}+\alpha_{c,g} \Delta_c,
\end{equation}
where 
\begin{itemize}
\item $p^{\mathrm{gen}}_{c,g}$ is the post-contingency active power produced by $g \in \Gcal$; 
\item $p^{\mathrm{gen}}_{0,g}$ is the base case active power of generator $g$; 
\item $\Delta_c$ represents the amount of post contingency redispatch; 
\item $\alpha_{c,g}$ denotes the weight of contribution to this redispatch by generator $g$. 
\end{itemize}
If a generator is operational but not selected to contribute to redispatch, it maintains its active power generation from the base case, i.e., 
\begin{align}
\alpha_{c,g}=0\quad\Rightarrow\quad p^{\mathrm{gen}}_{c,g}=p^{\mathrm{gen}}_{0,g}.
\end{align}

In the presence of post-contingency generator active power limits (i.e., when the assumption (\ref{111}) does not hold), a generator that contributes to a given contingency can readjust its active power generation to the extent that its capacity limits permit. This behavior can be formulated using logical functions, and Min-Max operators. For every $ c \in \Ccal$ and $g \in \Gcal$, the logical functions representation of generator active power contingency response, using disjunction of linear constraints, is as follows:
 \begin{subequations}\label{eq:activep}
\begin{align} 
& p^{\mathrm{gen}}_{c,g} = p^{\max}_{c,g},&&\hspace{-0.2mm}\mathrm{if}\qquad p^{\mathrm{max}}_{c,g}< p^{\mathrm{gen}}_{0,g}+\alpha_{c,g} \Delta_c, 
\label{eq:activep2} \\
& p^{\mathrm{gen}}_{c,g} = p^{\min}_{c,g},&&\hspace{-0.2mm}\mathrm{if}\qquad p^{\mathrm{min}}_{c,g}> p^{\mathrm{gen}}_{0,g}+\alpha_{c,g} \Delta_c, 
\label{eq:activep3}\\
&p^{\mathrm{gen}}_{c,g}=p^{\mathrm{gen}}_{0,g}+\alpha_{c,g} \Delta_c ,&&\hspace{-6.5mm} \qquad \mathrm{otherwise} 
. \label{eq:activep1}
\end{align}
\end{subequations}

Another way of formulating (\ref{eq:activep}) is by introducing two binary variables $x^{P^+}_{c,g},\, x^{P^-}_{c,g} \in \{0,1\}$ for every $c\in\mathcal{C}$ and $g\in\mathcal{G}$,  to indicate the mode of operation. Then, the three conditions in (\ref{eq:activep}) can be stated as
\begin{subequations}\label{eq:mixa1}
\begin{align}
&\!\!\!p^{\mathrm{gen}}_{0,g}\!+\!\alpha_{c,g}\Delta_c \!-\! p^{\mathrm{gen}}_{c,g} \leq M_{c,g}(1\!-\!x^{P^+}_{c,g})\; &&\hspace{0mm}\forall c\in\mathcal{C}, \; g\in\mathcal{G},\!\! \label{eq:mixa11}\\
&\!\!\!p^{\mathrm{gen}}_{c,g}\!-\!p^{\mathrm{gen}}_{0,g}\!-\!\alpha_{c,g}\Delta_c \leq M_{c,g}(1\!-\!x^{P^-}_{c,g})\; &&\hspace{0mm}\forall c\in\mathcal{C}, \; g\in\mathcal{G},\!\! \label{eq:mixa12}\\
&\!\!\!p^{\mathrm{max}}_{c,g}-p^{\mathrm{gen}}_{c,g} \leq (p^{\mathrm{max}}_{c,g}-p^{\mathrm{min}}_{c,g})\,x^{P^+}_{c,g}\; &&\hspace{0mm}\forall c\in\mathcal{C}, \; g\in\mathcal{G},\!\! \label{eq:mixa13}\\
&\!\!\!p^{\mathrm{gen}}_{c,g}-p^{\mathrm{min}}_{c,g} \leq (p^{\mathrm{max}}_{c,g}-p^{\mathrm{min}}_{c,g})\,x^{P^-}_{c,g}\; &&\hspace{0mm}\forall c\in\mathcal{C}, \; g\in\mathcal{G},\!\! \label{eq:mixa14}
\end{align}
\end{subequations}
where the ``big-M'' multipliers, $M_{c,g}$, are chosen sufficiently large to ensure that (\ref{eq:mixa1}) is equivalent to (\ref{eq:activep}).
Observe that according to (\ref{eq:mixa11}) and (\ref{eq:mixa12})
\begin{align}
&x^{P^+}_{c,g} = x^{P^-}_{c,g}=1 \;\;\Rightarrow\;\; p^{\mathrm{min}}_{c,g}\leq p^{\mathrm{gen}}_{c,g} = p^{\mathrm{gen}}_{0,g}+\alpha_{c,g}\Delta_c\leq p^{\mathrm{max}}_{c,g}\nonumber\\
&x^{P^+}_{c,g}=1,\; x^{P^-}_{c,g}=0 \;\;\Rightarrow\;\; p^{\mathrm{max}}_{c,g}=p^{\mathrm{gen}}_{c,g}\leq p^{\mathrm{gen}}_{0,g}+\alpha_{c,g}\Delta_c\nonumber\\
&x^{P^+}_{c,g}=0,\; x^{P^-}_{c,g}=1 \;\;\Rightarrow\;\; p^{\mathrm{gen}}_{0,g}+\alpha_{c,g}\Delta_c\leq p^{\mathrm{gen}}_{c,g} = p^{\mathrm{min}}_{c,g}\nonumber\\
&x^{P^+}_{c,g} = x^{P^-}_{c,g}=0 \;\;\Rightarrow\;\; p^{\mathrm{min}}_{c,g}= p^{\mathrm{gen}}_{c,g} = p^{\mathrm{max}}_{c,g}.\nonumber
\end{align}

\begin{figure}[t]
	\centering
	\includegraphics[width=\columnwidth]{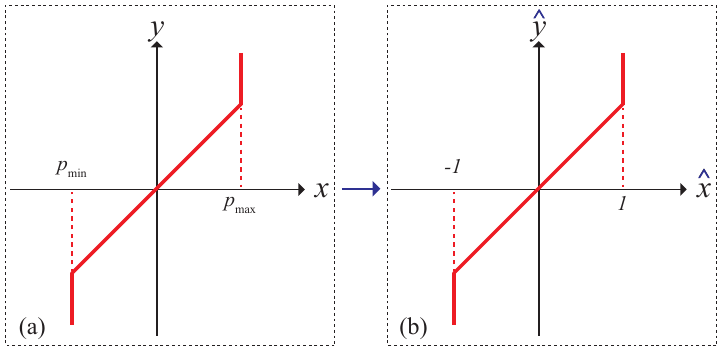}
	\caption{Generator active power response relation between pre- and post-contingency: (a) Actual characteristics: $x=p^{\mathrm{gen}}_{c,g}$, $y=p^{\mathrm{gen}}_{0,g}+\alpha_{c,g}\Delta_c$, (b) Normalization of the actual characteristics: $\hat{x}=\frac{2p^{\mathrm{gen}}_{c,g}-p^{\max}_{c,g}+p^{\min}_{c,g}}{p^{\max}_{c,g}+p^{\min}_{c,g}}$, $\hat{y}=\frac{2(p^{\mathrm{gen}}_{0,g}+\alpha_{c,g}\Delta_c)-p^{\max}_{c,g}+p^{\min}_{c,g}} {p^{\max}_{c,g}+p^{\min}_{c,g}}$.}
	\label{fig:fig1}
\end{figure}

Additionally, 
generator active power contingency response can be expressed using Min and Max operators as
\begin{align}
\!\!\!\pbf^{\mathrm{gen}}_c\!=\!\max\{\pbf_c^{\min}\!,\min\{\Cbf_{\!c}(\pbf^{\mathrm{gen}}_0\!+\!\boldsymbol{\alpha}_c\Delta_c),\pbf^{\max}_c\}\} \;\;\; \forall c\!\in\!\mathcal{C},\!\!\! \label{eq:minmaxa}
\end{align}
which can be equivalently formulated based on the defined set~(\ref{eq:set_orig}), for every $c \in \Ccal$ and $g \in \Gcal$, as
{\small
\begin{align} 
\!\!\!\!\Big(\frac{
	2p^{\mathrm{gen}}_{c,g}-\!p^{\max}_{c,g}\!+\!p^{\min}_{c,g}
}{
	p^{\max}_{c,g}+p^{\min}_{c,g}},\frac{2(p^{\mathrm{gen}}_{0,g}\!+\!\alpha_{c,g}\Delta_c)\!-p^{\max}_{c,g}\!+\!p^{\min}_{c,g}} {p^{\max}_{c,g}+p^{\min}_{c,g}
}\Big)
\!\in\! \Fcal_{\frac{\pi}{4}}.\!\!\!\! \label{eq:minmaxa1}
\end{align}}
\noindent \!\!\!\! Equations (\ref{eq:activep}) - (\ref{eq:minmaxa1}) are all equivalent and imply a piecewise-smooth model as shown in Figure~\ref{fig:fig1} (a).

 \subsection{Generator Reactive Power Contingency Response}
Ideally, the generators that contribute to a given contingency redispatch may need to readjust their post-contingency reactive power production in order to retain the base case voltage magnitude,
\begin{align}
|v_c|=|v_0|.
\end{align}
However, in the presence of tight reactive power limits, maintaining pre-contingency voltage magnitudes might not be possible. Therein, 
\begin{itemize}
\item the bus voltage magnitude can drop below its base-case level if all generator reactive power upper limits of the bus are binding;
\item  the bus voltage magnitude is allowed to rise above its base-case level if all generator reactive power lower limits of the bus are binding.
\end{itemize}
This requirement can be formulated as follows: 
\begin{subequations}\label{eq:ractivep}
\begin{align}
&\!\!\! q^{\min}_{c,g} \!\leq\! q^{\mathrm{gen}}_{c,g} \!\leq\! q^{\max}_{c,g} ,&&\hspace{-1mm}|v_{c,i}|\!=\!|v_{0,i}|&&&&\hspace{-4mm} \forall c \!\in\! \Ccal, g \!\in\! \Gcal, i \!\in\! \Ncal,\!\!\!\! \label{eq:ractivep1}\\
&\!\!\! q^{\mathrm{gen}}_{c,g} = q^{\max}_{c,g} ,&&\hspace{-1mm}|v_{c,i}|\!<\!|v_{0,i}|&&&&\hspace{-4mm} \forall c \!\in\! \Ccal, g \!\in\! \Gcal, i \!\in\! \Ncal,\!\!\!\! \label{eq:ractivep2} \\
&\!\!\! q^{\mathrm{gen}}_{c,g} = q^{\min}_{c,g} ,&&\hspace{-1mm}|v_{c,i}|\!>\!|v_{0,i}|&&&&\hspace{-4mm} \forall c \!\in\! \Ccal, g \!\in\! \Gcal, i \!\in\! \Ncal.\!\!\!\!\label{eq:ractivep3}
\end{align}
\end{subequations}

Alternatively, we can reformulate (\ref{eq:ractivep}) with respect to auxiliary binary variables: 
\begin{subequations}\label{eq:mixr1}
\begin{align}
&|v_{0,i}|\!-\!|v_{c,i}| \leq (v^{\mathrm{max}}_{i}\!-\!v^{\mathrm{min}}_{i})(1\!-\!x^{Q^+}_{c,g}) 
&&\hspace{0mm}\forall c\!\in\!\mathcal{C},  g\!\in\!\mathcal{G},\!\! \label{eq:mixr11}\\
&|v_{c,i}|\!-\!|v_{0,i}| \leq (v^{\mathrm{max}}_{i}\!-\!v^{\mathrm{min}}_{i})(1\!-\!x^{Q^-}_{c,g}) 
&&\hspace{0mm}\forall c\!\in\!\mathcal{C},  g\!\in\!\mathcal{G},\!\! \label{eq:mixr12}\\
&q^{\mathrm{max}}_{c,g}\!-\!q^{\mathrm{gen}}_{c,g} \leq (q^{\mathrm{max}}_{c,g}\!-\!q^{\mathrm{min}}_{c,g})x^{Q^+}_{c,g}\!\!\!\! &&\hspace{0mm}\forall c\!\in\!\mathcal{C}, \, g\!\in\!\mathcal{G},\!\! \label{eq:mixr13}\\
&q^{\mathrm{gen}}_{c,g}\!-\!q^{\mathrm{min}}_{c,g} \leq (q^{\mathrm{max}}_{c,g}\!-\!q^{\mathrm{min}}_{c,g})x^{Q^-}_{c,g}\!\!\!\! &&\hspace{0mm}\forall c\!\in\!\mathcal{C}, \, g\!\in\!\mathcal{G}.\!\! \label{eq:mixr14}
\end{align}
\end{subequations}
where $i\in\mathcal{N}$ is the bus where $g$ is located and $x^{Q^+}_{c,g},\, x^{Q^-}_{c,g} \in \{0,1\}$ denote the introduced binary variables. Observe that
\begin{align}
&x^{Q^+}_{c,g}= x^{Q^-}_{c,g} = 1 \;\Rightarrow\;|v_{c,i}|=|v_{0,i}|\;\wedge\;
q^{\mathrm{min}}_{c,g}\leq q^{\mathrm{gen}}_{c,g}\leq q^{\mathrm{max}}_{c,g},\nonumber\\
&x^{Q^+}_{c,g}=0,\;\; x^{Q^-}_{c,g} = 1 \;\Rightarrow\;|v_{c,i}|\leq|v_{0,i}|\;\wedge\;
q^{\mathrm{gen}}_{c,g}= q^{\mathrm{max}}_{c,g},\nonumber\\
&x^{Q^+}_{c,g}=1,\;\; x^{Q^-}_{c,g} = 0 \;\Rightarrow\;|v_{c,i}|\geq|v_{0,i}|\;\wedge\;
q^{\mathrm{gen}}_{c,g}= q^{\mathrm{min}}_{c,g},\nonumber\\
&x^{Q^+}_{c,g}= x^{Q^-}_{c,g} = 0 \;\Rightarrow\;
q^{\mathrm{min}}_{c,g}= q^{\mathrm{gen}}_{c,g}= q^{\mathrm{max}}_{c,g}.\nonumber
\end{align}


In addition to the representation in (\ref{eq:ractivep}) and (\ref{eq:mixr1}), for every $c\in\Ccal$,
generator reactive power contingency response can be expressed using Min and Max operators as follows:
\begin{align}
 &\!\!\!\!\!\!\min\!\big\{\!\max\{0,\Gbf_c(|\vbf_0|\!-\!|\vbf_c|)\},\max\{0,\qbf^{\max}_c\!-\!\qbf^{\mathrm{gen}}_c\}\big\}\!=\!\nonumber\\
&\!\!\!\!\!\!\min\!\big\{\!\max\{0,\Gbf_c(|\vbf_c|\!-\!|\vbf_0|)\},\max\{0,\qbf^{\mathrm{gen}}_c\!-\!\qbf^{\min}_c\}\big\}\!=\!0. \!\!\!\!
\end{align}
Equation (\ref{eq:ractivep})  can be equivalently formulated based on the defined set (\ref{eq:set_orig}) for every $c \in \Ccal$, $g \in \Gcal$ and $i \in \Ncal$ as
\begin{align} 
\Big(\frac{2q^{\mathrm{gen}}_{c,g}-q^{\max}_{c,g}-q^{\min}_{c,g}}
{q^{\max}_{c,g}-q^{\min}_{c,g}},\;\; |v_{0,i}|-|v_{c,i}|\Big) \in \Fcal_{0}.\label{eq:minmaxr1}
\end{align}

\begin{figure}[h]
	\centering 
	\includegraphics[width=\columnwidth]{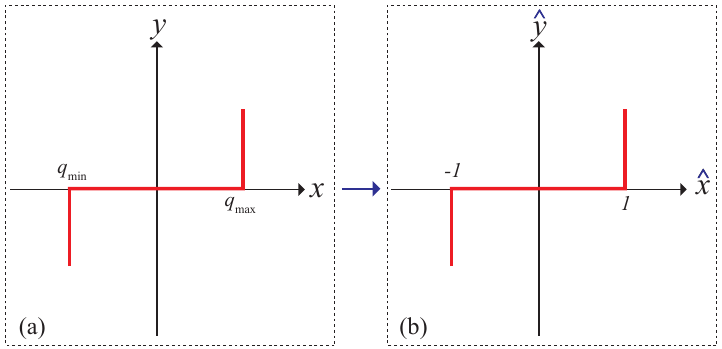}
	\caption{Generator reactive power response relation between pre- and post-contingency: (a) Actual characteristics: $x=q^{\mathrm{gen}}_{c,g}$, $y=|v_{0,i}|-|v_{c,i}| $, (b) Normalization of the actual characteristics: $\hat{x}=\frac{2q^{\mathrm{gen}}_{c,g}-q^{\max}_{c,g}-q^{\min}_{c,g}}{q^{\max}_{c,g}-q^{\min}_{c,g}}$, $\hat{y}=|v_{0,i}|-|v_{c,i}|$.}
	\label{fig:fig2}
\end{figure}

The formulations, (\ref{eq:ractivep}) -- (\ref{eq:minmaxr1}), refer to the PV/PQ switching as demonstrated in Figure~\ref{fig:fig2} (a). Herein, PV bus means that the bus voltage magnitude and the generation level of active power are fixed whereas its voltage phase angle and reactive power generation are varying. When it hits the reactive power capacity limits to sustain the value of voltage magnitude in the base case, the bus type needs to become PQ. This bus type means that the power generations, e.g., real and reactive, are constant, whereas the voltage magnitude and phase angle are varying. Similar to the characteristic of active power, reactive power contingency response implies a piecewise-smooth  model. 

\section{Security-constrained Optimal Power Flow}

This section elaborates the active and reactive power contingency response models. Herein, devised objective function minimizes the generational cost in the base case and secures the grid to contingency scenarios.

\subsection{Non-differentiable Contingency Response in SCOPF Formulation}

SCOPF with respect to the outage of an arbitrary set of generators and lines can be formulated as 
{\small{
\begin{subequations}\label{eq:OPFLC}
	\begin{align}
	&\hspace{-0.0mm} \underset{
	}{\text{~~~minimize~~~}}
	&&\hspace{-0.5cm}  h(\pbf^{\mathrm{gen}}_0) 
	\hspace{-5mm}   \hspace{0mm}\label{eq:OPFLC_obj}\\ \nonumber \\
	&\hspace{-0.0mm} \text{~~~subject to~~~~~~~} && &&&&\nonumber\\
	&&&\hspace{-2.25cm}\sbf^{\mathrm{dem}}_c+\mathrm{diag}\{\!\vbf_c\vbf_c^{\ast}\,\Ybf^{\ast}_{\!c}\} 
	=\Gbf^{\top}_c(\pbf^{\mathrm{gen}}_c+i \qbf^{\mathrm{gen}}_c) &&&& \hspace{-2.5cm}\forall c\in\mathcal{C}\hspace{0mm} \label{eq:OPFLC_cons_4}\hspace{0mm}\\
	&\hspace{-0.0mm}&&\hspace{-2.25cm}|\mathrm{diag}\{\vec{\Lbf}_c^{\phantom{\ast}}\;\!\vbf_c\vbf_c^{\ast}\,\vec{\Ybf}^{\ast}_{\!c}\}|\leq\sbf^{\mathrm{max}}_c &&&& \hspace{-2.5cm}\forall c\in\mathcal{C}\hspace{0mm} \label{eq:OPFLC_cons_5}\\
	&\hspace{-0.0mm}&&\hspace{-2.25cm}|\mathrm{diag}\{\cev{\Lbf}_c^{\phantom{\ast}}\;\!\vbf_c\vbf_c^{\ast}\,\cev{\Ybf}^{\ast}_{\!c}\}|\leq\sbf^{\mathrm{max}}_c &&&& \hspace{-2.5cm}\forall c\in\mathcal{C}\hspace{0mm} \label{eq:OPFLC_cons_6}\\
	&\hspace{-0.0mm}&&\hspace{-2.25cm}\pbf^{\min}_c\leq \pbf^{\mathrm{gen}}_c \leq \pbf^{\max}_c &&&& \hspace{-2.5cm}\forall c\in\mathcal{C}\hspace{0mm}  \label{eq:OPFLC_cons_7}\\
	&\hspace{-0.0mm}&&\hspace{-2.25cm}\qbf^{\min}_c\leq \qbf^{\mathrm{gen}}_c \leq \qbf^{\max}_c &&&& \hspace{-2.5cm}\forall c\in\mathcal{C}\hspace{0mm}  \label{eq:OPFLC_cons_8}\\
	&\hspace{-3.0mm}&&\hspace{-2.25cm}\vbf^{\min}\leq|\vbf_c|\,\leq\vbf^{\max} &&&& \hspace{-2.5cm}\forall c\in\mathcal{C}\hspace{0mm}  \label{eq:OPFLC_cons_9}\\
	&\hspace{-0.0mm}&& &&&&\nonumber\\
	&\hspace{-0.0mm}&&\hspace{-2.3cm} \Big(\frac{2\pbf^{\mathrm{gen}}_c\!-\!\pbf^{\max}_c\!+\!\pbf^{\min}_c}{\pbf^{\max}_c\!+\!\pbf^{\min}_c},\frac{2\Cbf_{\!c}(\pbf^{\mathrm{gen}}_0\!+\!\boldsymbol{\alpha_c}\Delta_c)\!-\!\pbf^{\max}_c\!+\!\pbf^{\min}_c}
	{\pbf^{\max}_c+\pbf^{\min}_c}\Big) 
	\nonumber\\
	&\hspace{-0.0mm}&&\hspace{-2.3cm} &&&& \hspace{-3.6cm} 
	\in \Fcal_{\frac{\pi}{4}}\,\, \; \forall c\in\mathcal{C}\hspace{0mm} \label{eq:OPFLC_cons_10}\\
	&\hspace{-0.0mm}&&\hspace{-2.3cm}
\Big(\frac{2\qbf^{\mathrm{gen}}_c\!-\!\qbf^{\max}_c\!-\!\qbf^{\min}_c}
{\qbf^{\max}_c-\qbf^{\min}_c}, \Gbf_c(|\vbf_0|\!-\!|\vbf_c|)\Big) \in \Fcal_{0} &&&& \hspace{-2.5cm}\forall c\in\mathcal{C}\hspace{0mm} \label{eq:OPFLC_cons_11}\\
	\nonumber\\
	&\hspace{-0mm}\text{~~~variables}\hspace{-2.0cm} && \hspace{-1cm}  \pbf^{\mathrm{gen}}_0,\pbf^{\mathrm{gen}}_c,\qbf^{\mathrm{gen}}_c, \boldsymbol{\alpha_c} \in\mathbb{R}^{|\mathcal{G}|}; \;  \vbf_0, \vbf_c\in\mathbb{C}^{|\mathcal{N}|}; \; \Delta_c\in\mathbb{R} \nonumber
	\end{align}
\end{subequations}}}
\!\!\!where (\ref{eq:OPFLC_obj}) represents the cost of producing power in the base case. (\ref{eq:OPFLC_cons_4}) fulfills the apparent power balances in the network, whereas  (\ref{eq:OPFLC_cons_5})–(\ref{eq:OPFLC_cons_6}) 
enforce the apparent power flow limits over transmission lines. (\ref{eq:OPFLC_cons_7})–(\ref{eq:OPFLC_cons_8}) represent the generation boundaries for active and reactive power. The inequality (\ref{eq:OPFLC_cons_9}) limits the nodal voltage magnitude. Equation (\ref{eq:OPFLC_cons_10}) represents coupling constraints on generators that relate pre- and post-contingency active power dispatch, while (\ref{eq:OPFLC_cons_11}) stands for the coupling constraints that relate pre- and post-contingency reactive power dispatch with respect to generator capacity limits. It should be noted that  (\ref{eq:OPFLC_cons_10}) and (\ref{eq:OPFLC_cons_11}) are the normalized versions of (\ref{eq:activep}) and (\ref{eq:ractivep}), respectively. 

Constraints (\ref{eq:OPFLC_cons_10}) and (\ref{eq:OPFLC_cons_11}) are major sources of complexity which is due to non-smoothness of the curves $\Fcal_{0}$ and $\Fcal_{\frac{\pi}{4}}$. The main reason behind their normalizations is to conveniently substitute these curves with a smooth sigmoid surrogate. 
In the remainder of this section, we employ different families of continuously-differentiable models as alternatives for (\ref{eq:OPFLC_cons_10}) and (\ref{eq:OPFLC_cons_11}).
 
\subsection{Continuously-differentiable Contingency Response}

\begin{figure}[t]
	\centering
	\includegraphics[scale=0.6]{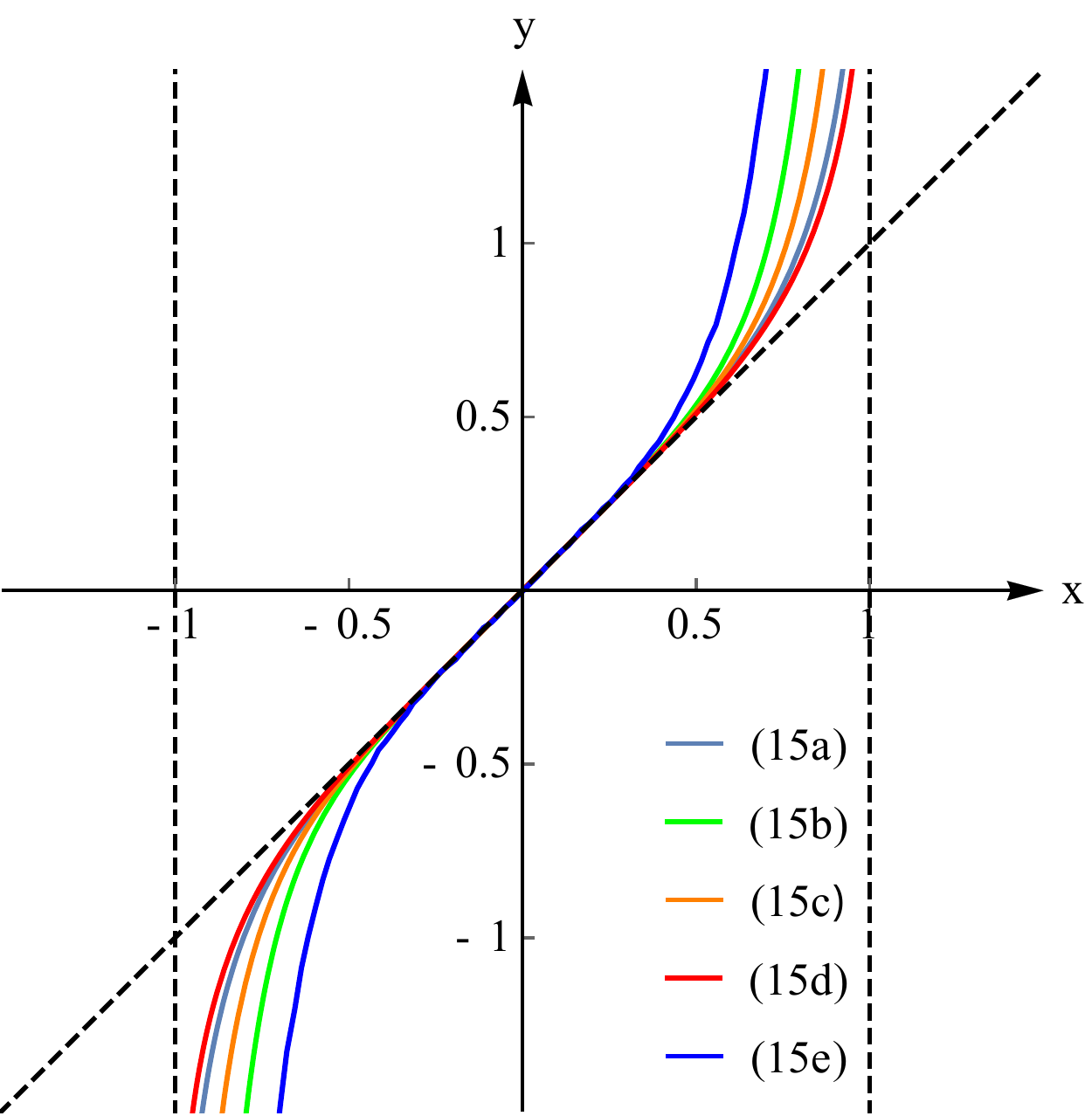}
	\caption{Sigmoid functions given in (\ref{eq:ac1})-(\ref{eq:ac5}) are normalized at the origin.}
	\label{fig:fig5_1}
\end{figure}

The set defined in (\ref{eq:set_orig}) represents a non-differentiable curve. To facilitate the task of local search algorithms, we define a family of continuously-differentiable curves that closely resemble generator contingency responses. 
Such surrogates can be constructed using any of the following sigmoid functions:
\begin{subequations}	\label{eq:14}
	\begin{align}
	&a_1(x)\triangleq\frac{\log(1+x)-\log(1-x)}{2} \label{eq:ac1}\\
	&a_2(x)\triangleq\frac{2}{\pi}\tan\left(\frac{\pi x}{2}\right)\label{eq:ac2}\\
	&a_3(x)\triangleq\frac{x}{\sqrt{1-x^2}}\label{eq:ac3}\\
	&a_4(x)\triangleq\frac{2}{\sqrt{\pi}}\mathrm{ierf}(x)\label{eq:ac4}\\
	&a_5(x)\triangleq\frac{x}{1-|x|}\label{eq:ac5}
	\end{align}
\end{subequations}
that are commonly-used in other areas.  These sigmoids are illustrated in Figure \ref{fig:fig5_1}.

\begin{figure*} 
	\centering
	\subfloat(a){%
		\includegraphics[width=0.17\linewidth]{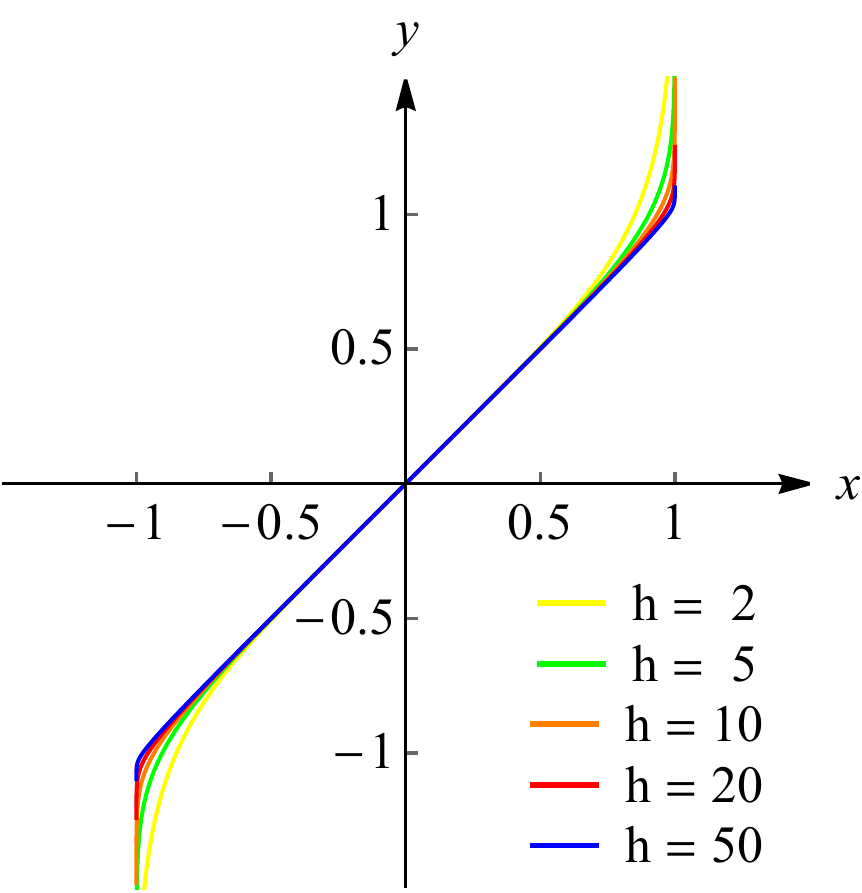}}
	\hfill
	\subfloat(b){%
		\includegraphics[width=0.17\linewidth]{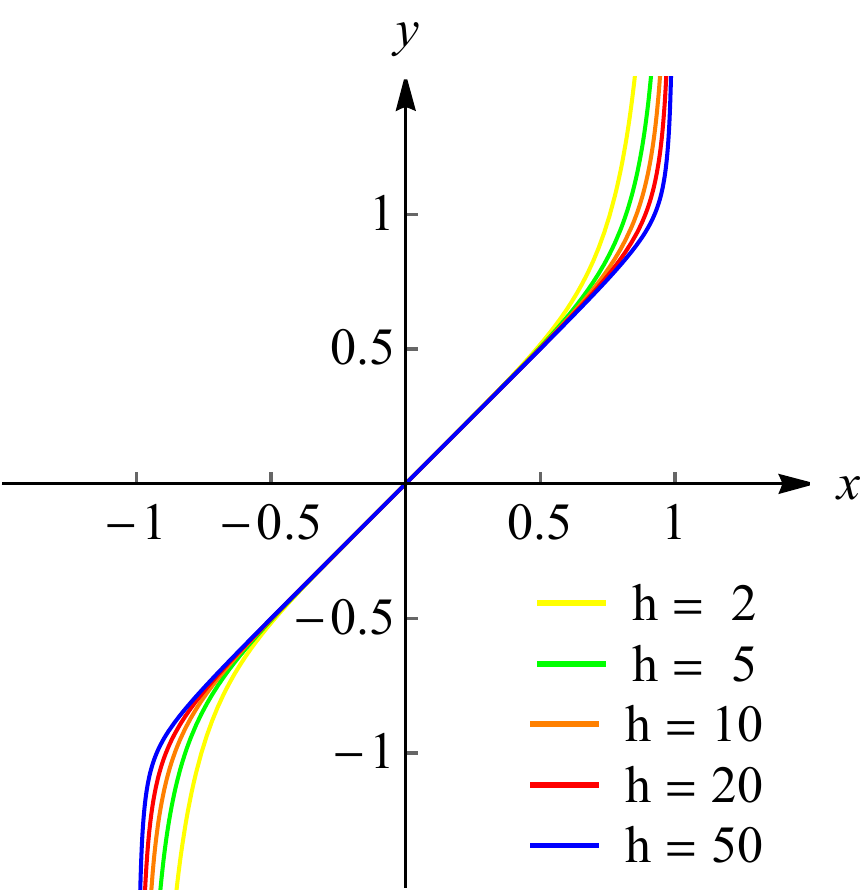}}
	\hfill
	\subfloat(c){%
	\includegraphics[width=0.17\linewidth]{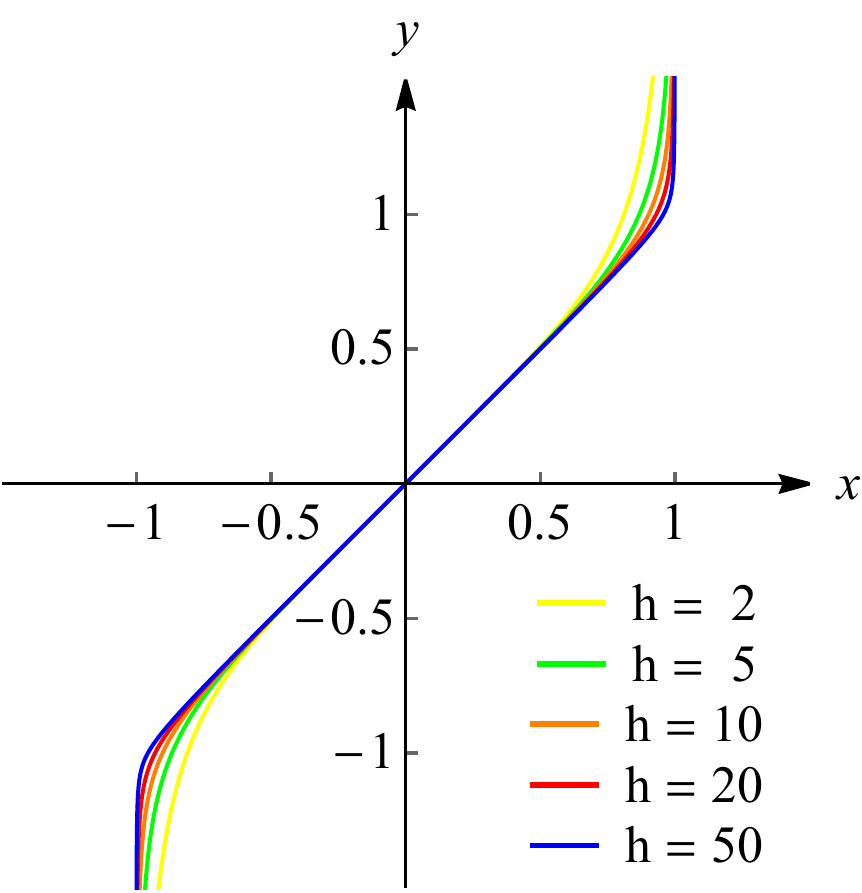}}
	\hfill
	\subfloat(d){%
	\includegraphics[width=0.17\linewidth]{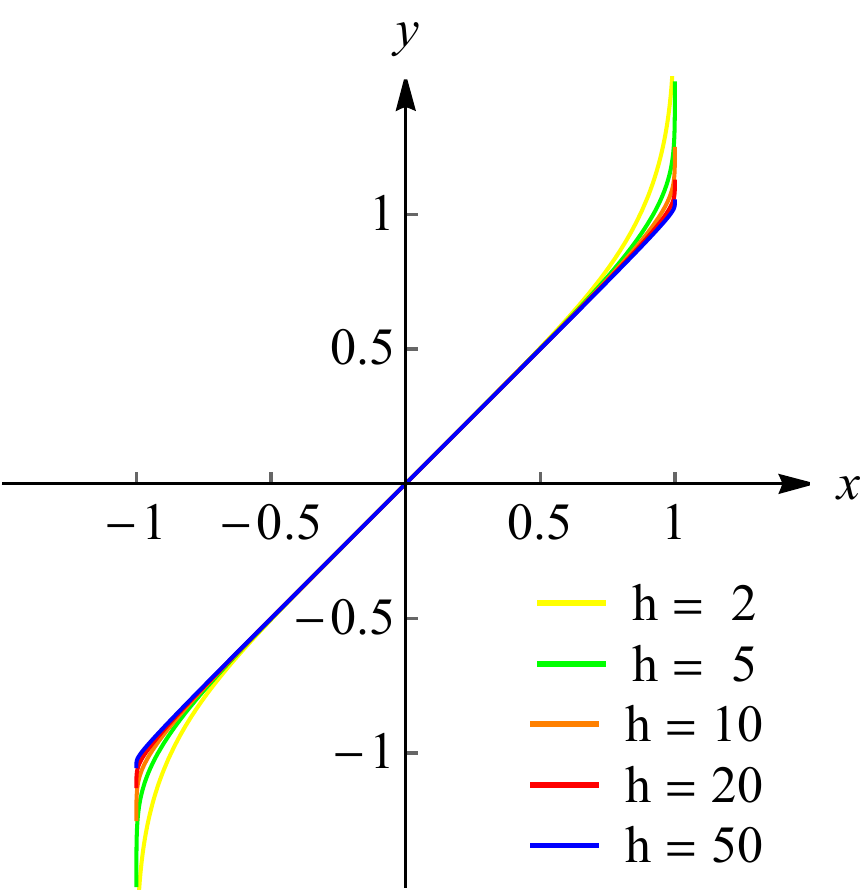}}
	\hfill
	\subfloat(e){%
	\includegraphics[width=0.17\linewidth]{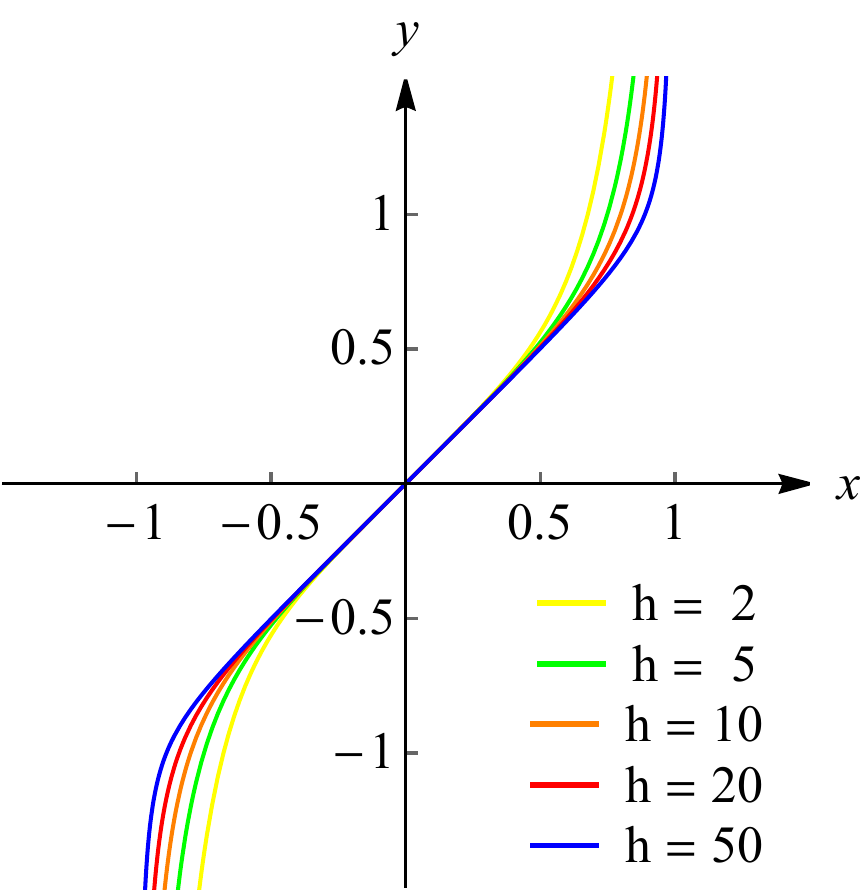}}
	\caption{The characteristics of $(x,y) \in \Fcal^{\mathrm{smooth}}_{\pi/4}$ for different $h$ values with the candidate curves: (a) (\ref{eq:ac1}), (b) (\ref{eq:ac2}), (c) (\ref{eq:ac3}), (d) (\ref{eq:ac4}), (e) (\ref{eq:ac5}).}
	\label{fig:fig5} 
\end{figure*}

\begin{figure*} 
	\centering
	\subfloat(a){%
		\includegraphics[width=0.17\linewidth]{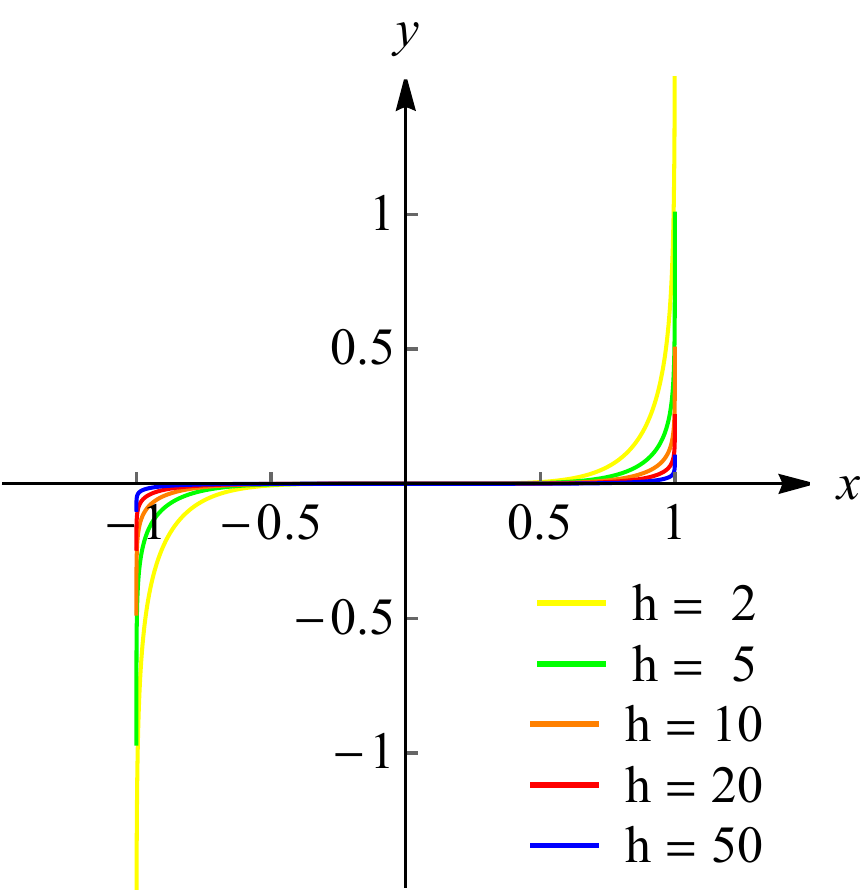}}
	\hfill
	\subfloat(b){%
		\includegraphics[width=0.17\linewidth]{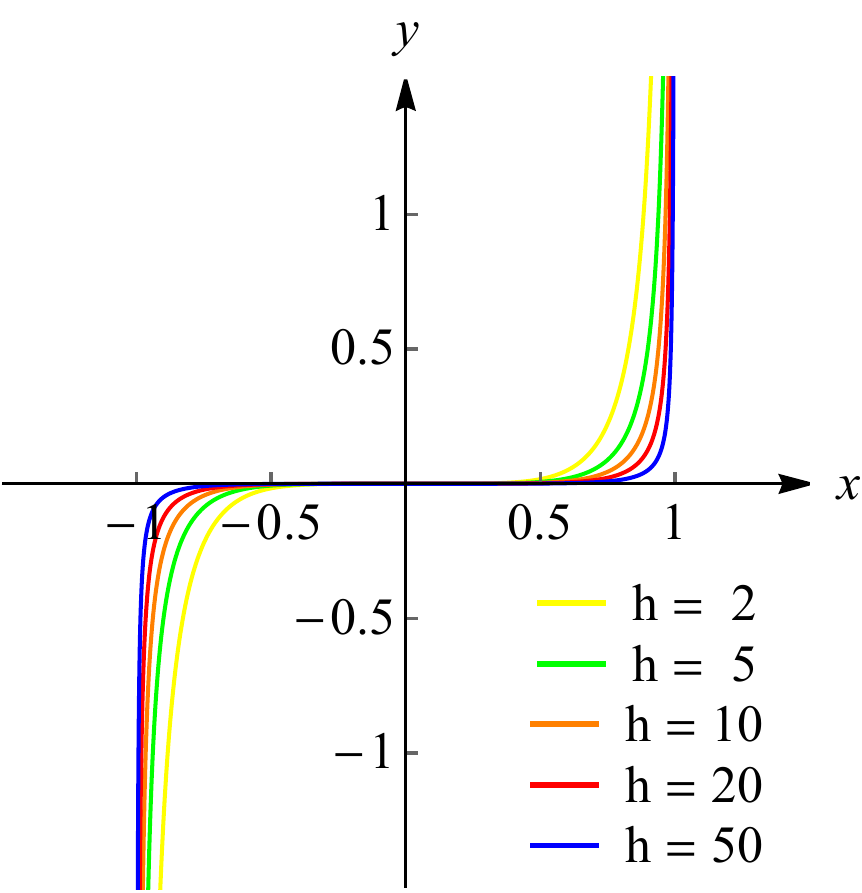}}
	\hfill
	\subfloat(c){%
		\includegraphics[width=0.17\linewidth]{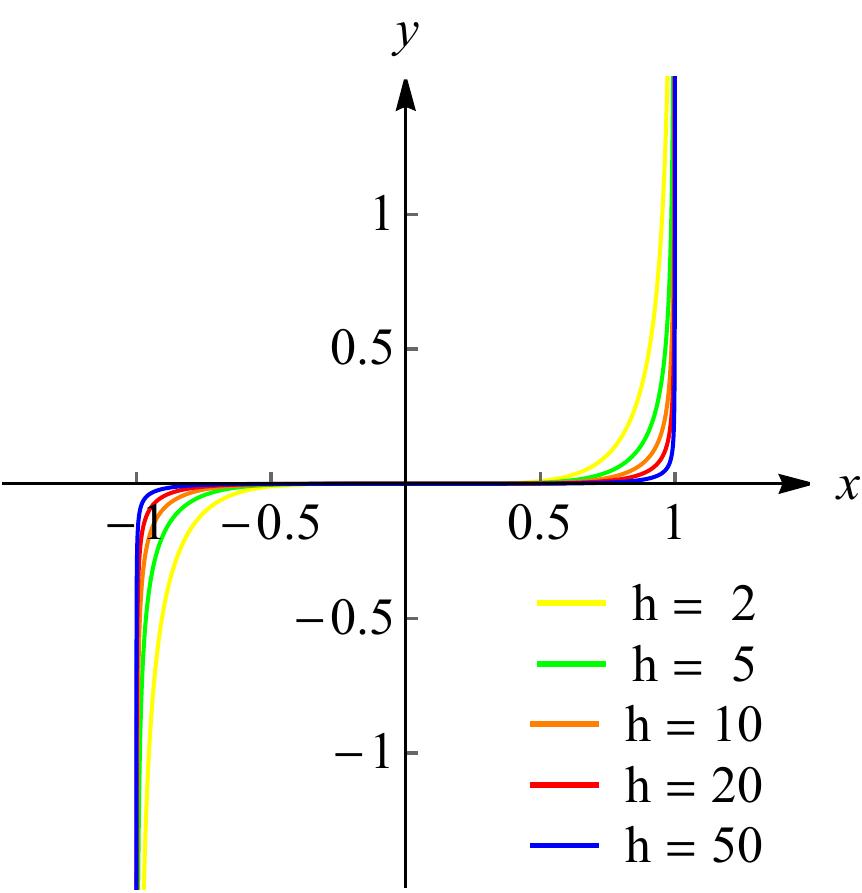}}
	\hfill
	\subfloat(d){%
		\includegraphics[width=0.17\linewidth]{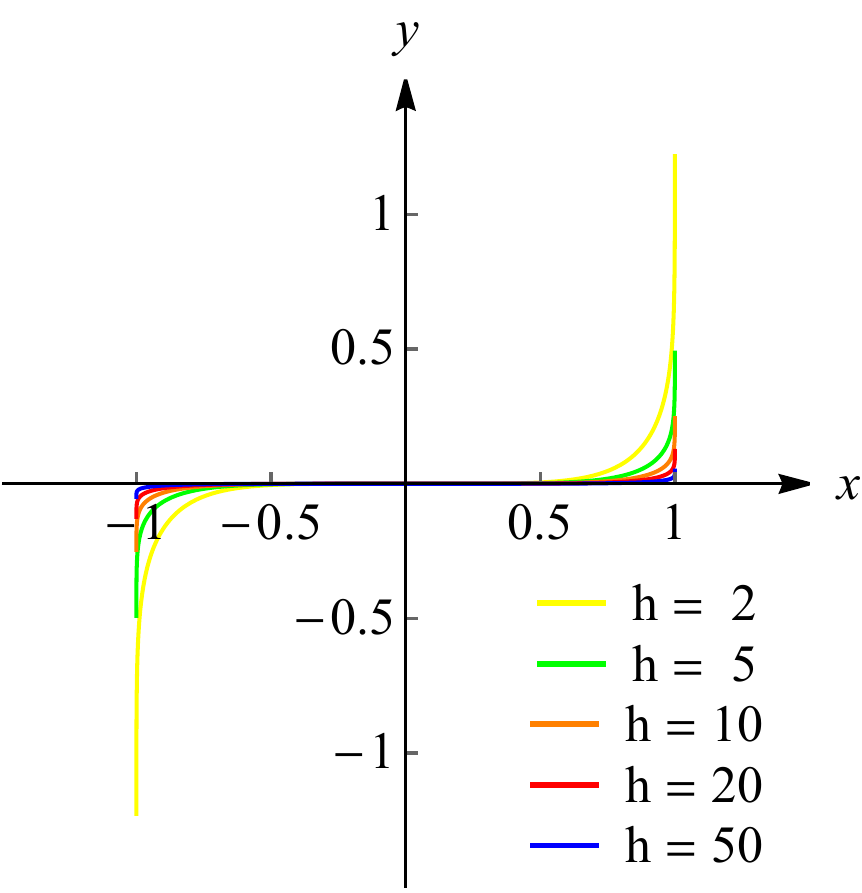}}
	\hfill
	\subfloat(e){%
		\includegraphics[width=0.17\linewidth]{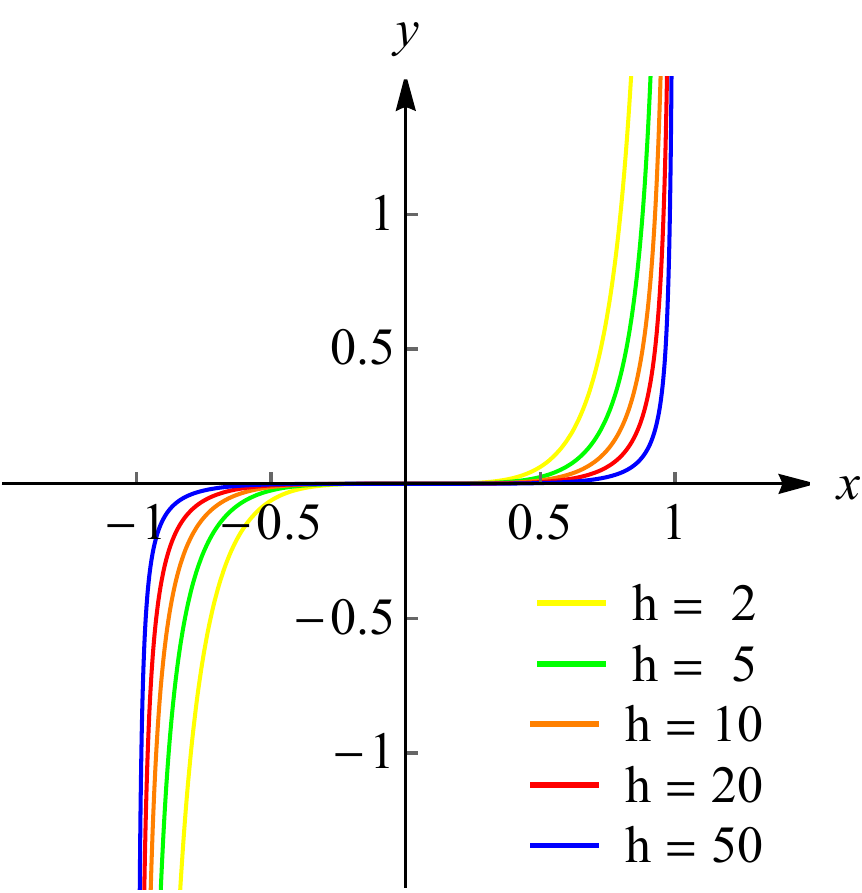}}
	\caption{The characteristics of $(x,y) \in \Fcal^{\mathrm{smooth}}_{0}$ for different $h$ values with the candidate curves: (a) (\ref{eq:ac1}), (b) (\ref{eq:ac2}), (c) (\ref{eq:ac3}), (d) (\ref{eq:ac4}), (e) (\ref{eq:ac5}).}
	\label{fig:fig6} 
\end{figure*}

\begin{figure*} 
	\centering
	\subfloat(a){%
		\includegraphics[width=0.17\linewidth]{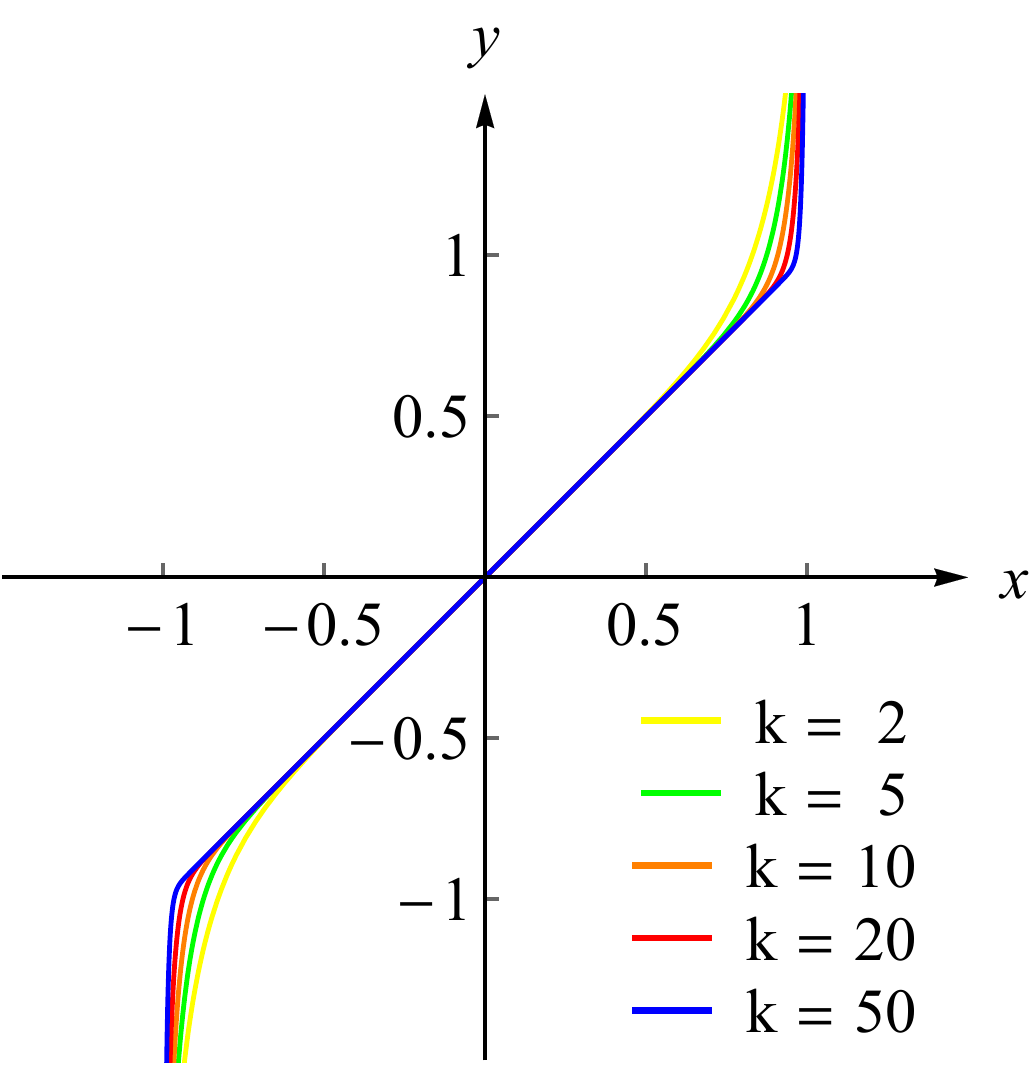}}
	\hfill
	\subfloat(b){%
		\includegraphics[width=0.17\linewidth]{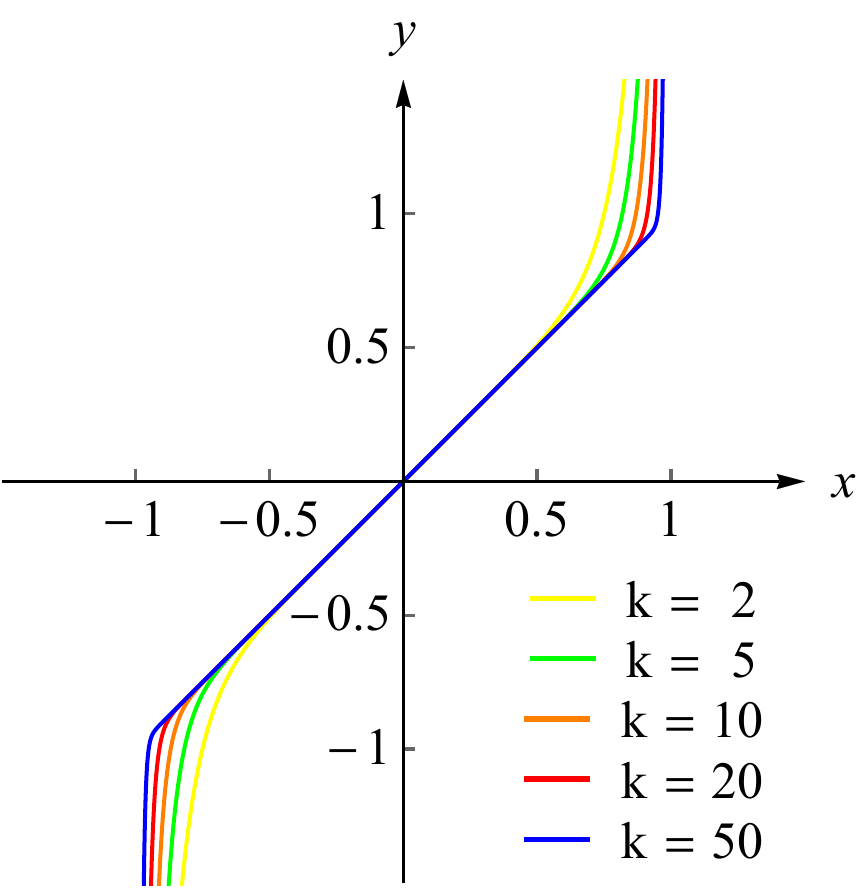}}
	\hfill
	\subfloat(c){%
		\includegraphics[width=0.17\linewidth]{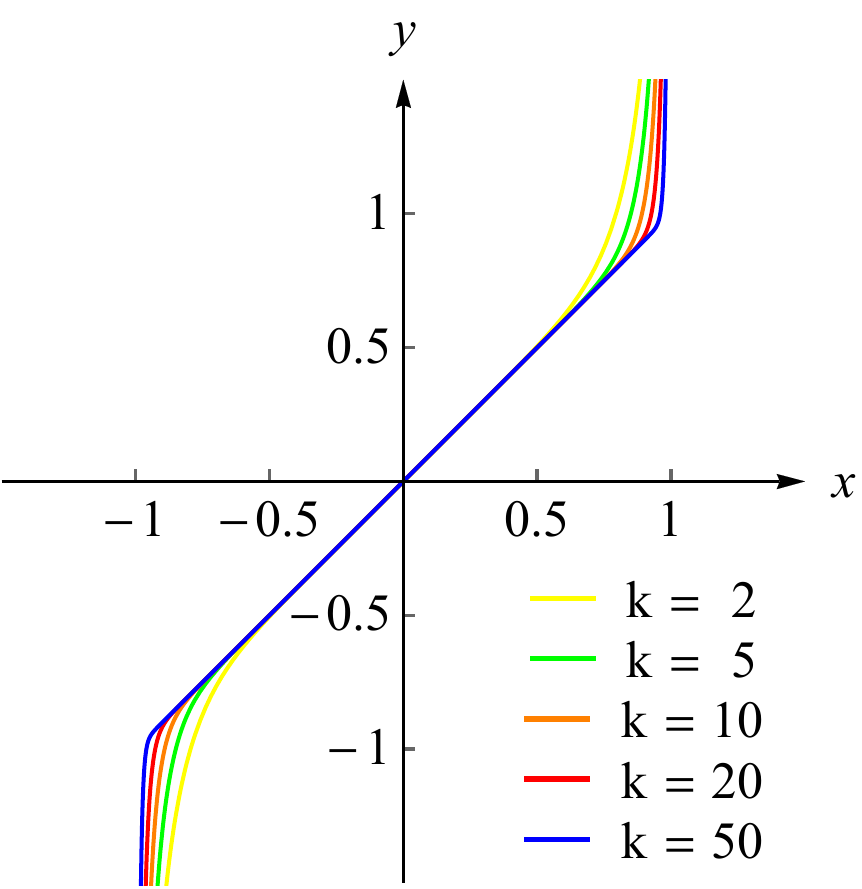}}
	\hfill
	\subfloat(d){%
		\includegraphics[width=0.17\linewidth]{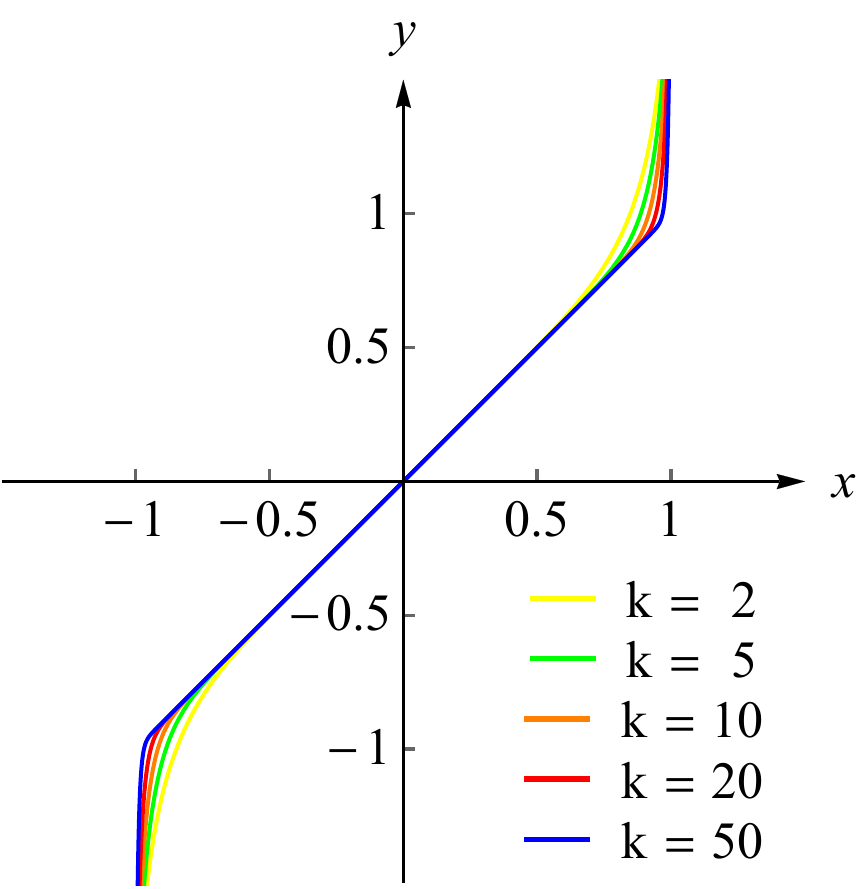}}
	\hfill
	\subfloat(e){%
		\includegraphics[width=0.17\linewidth]{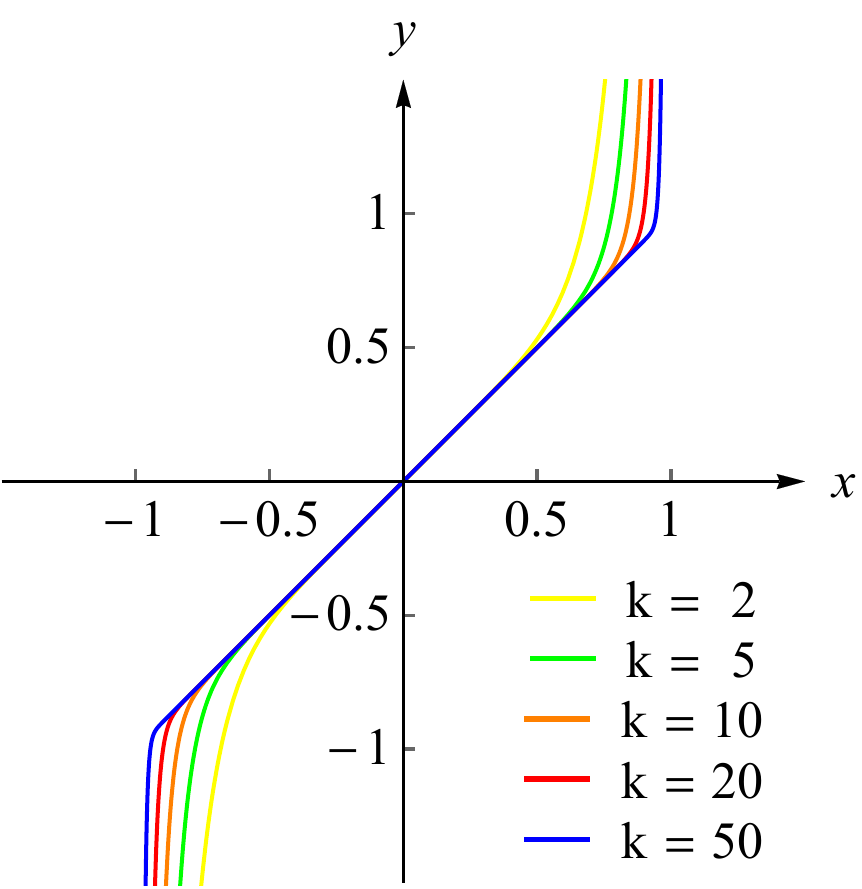}}
	\caption{The characteristics of $(x,y) \in \Fcal^{\mathrm{smooth}}_{\pi/4}$ for different $k$ values with the candidate curves: (a) (\ref{eq:ac1}), (b) (\ref{eq:ac2}), (c) (\ref{eq:ac3}), (d) (\ref{eq:ac4}), (e) (\ref{eq:ac5}).}
	\label{fig:fig7} 
\end{figure*}

\begin{figure*} 
	\centering
	\subfloat(a){%
		\includegraphics[width=0.17\linewidth]{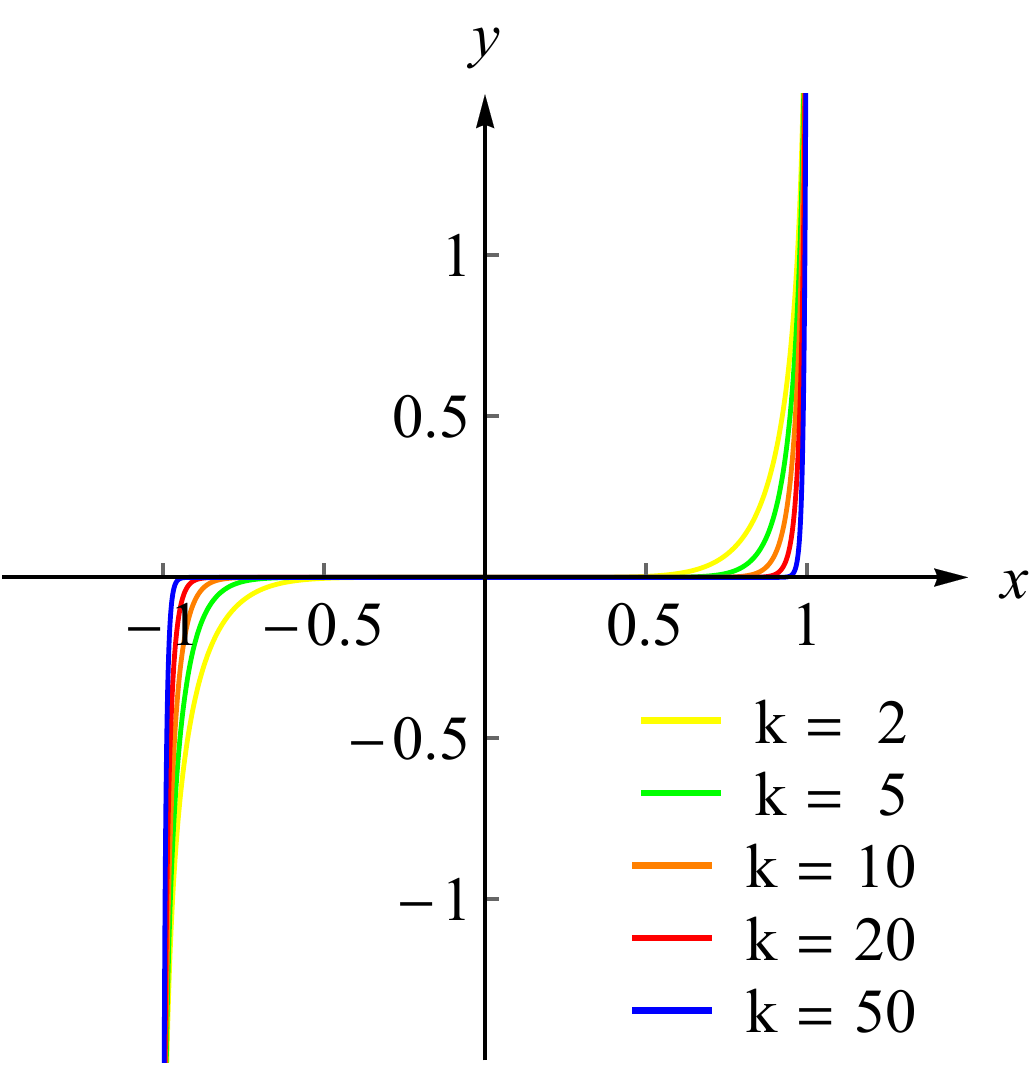}}
	\hfill
	\subfloat(b){%
		\includegraphics[width=0.17\linewidth]{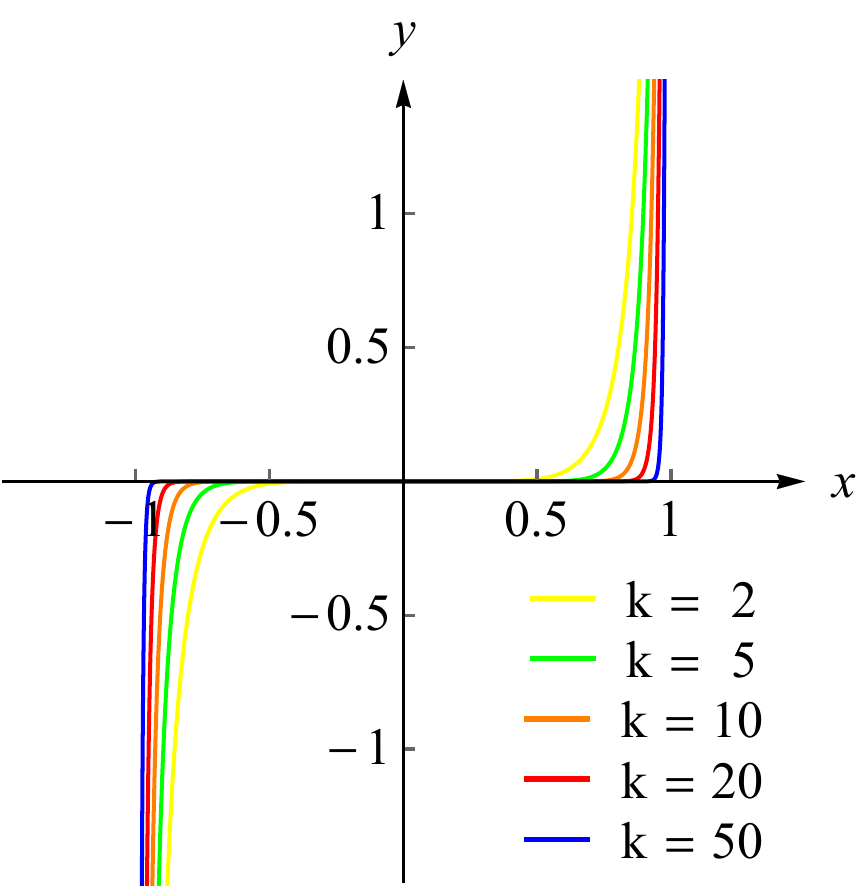}}
	\hfill
	\subfloat(c){%
		\includegraphics[width=0.17\linewidth]{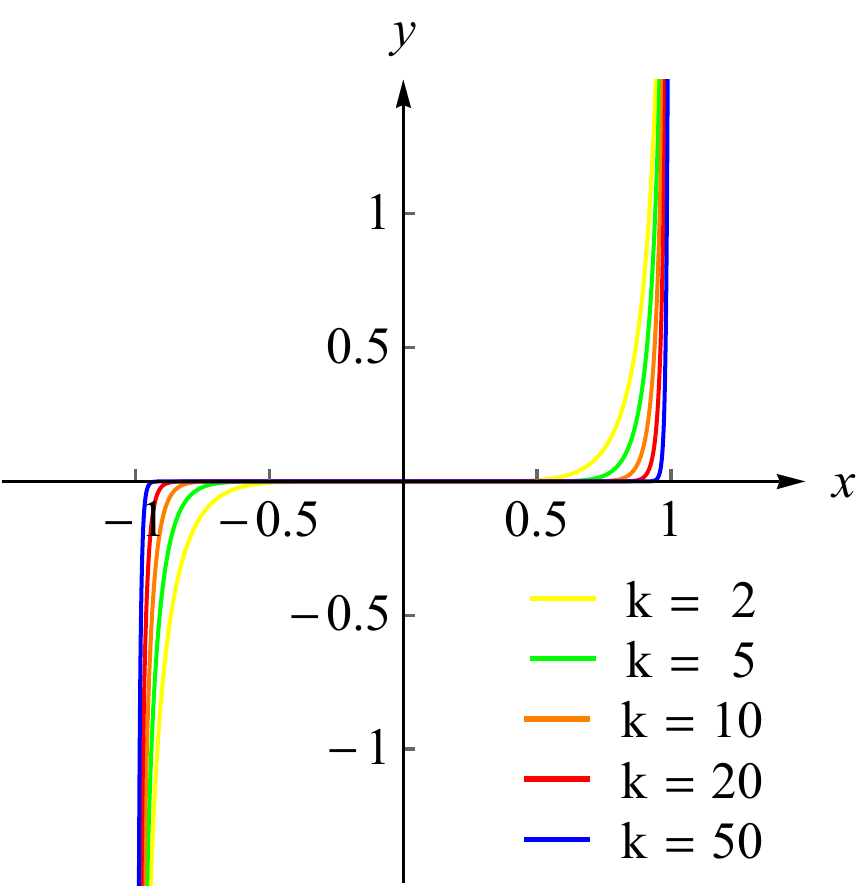}}
	\hfill
	\subfloat(d){%
		\includegraphics[width=0.17\linewidth]{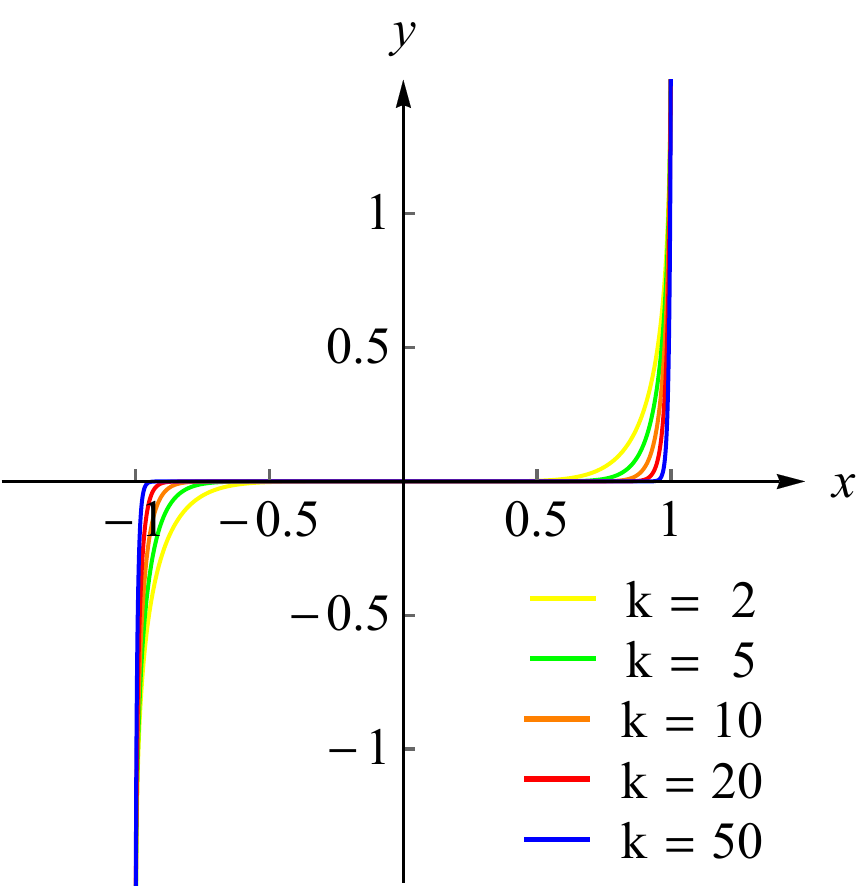}}
	\hfill
	\subfloat(e){%
		\includegraphics[width=0.17\linewidth]{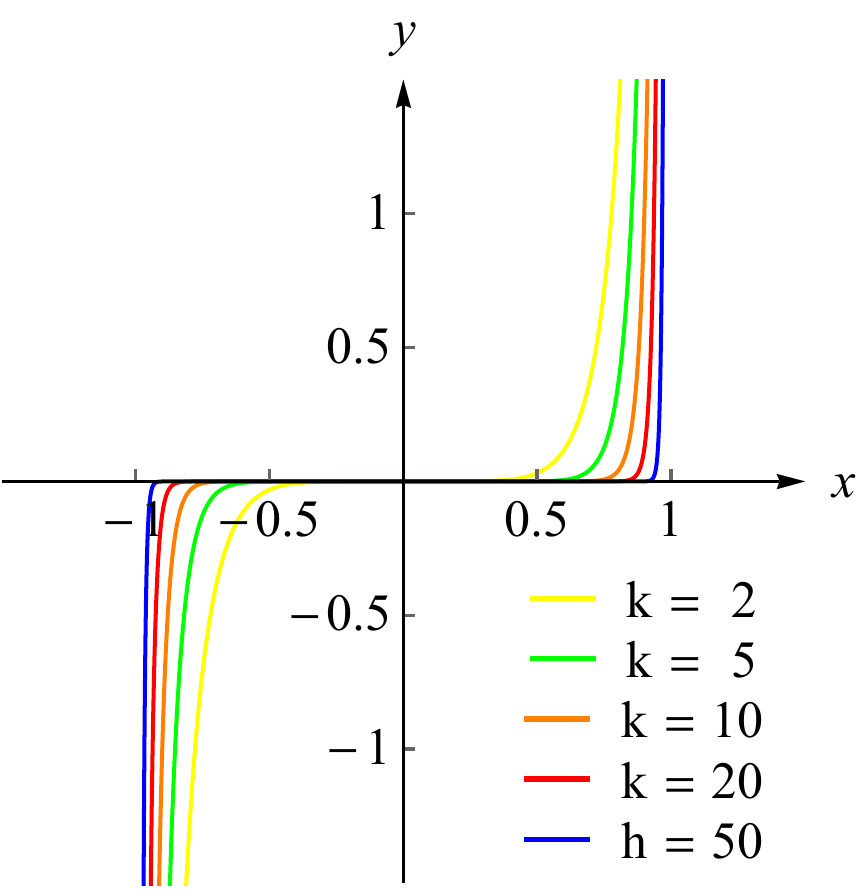}}
	\caption{The characteristics of $(x,y) \in \Fcal^{\mathrm{smooth}}_{0}$ for different $k$ values with the candidate curves: (a) (\ref{eq:ac1}), (b) (\ref{eq:ac2}), (c) (\ref{eq:ac3}), (d) (\ref{eq:ac4}), (e) (\ref{eq:ac5}).}
	\label{fig:fig8} 
\end{figure*}

\begin{figure*} 
	\centering
	\subfloat(a){%
		\includegraphics[width=0.17\linewidth]{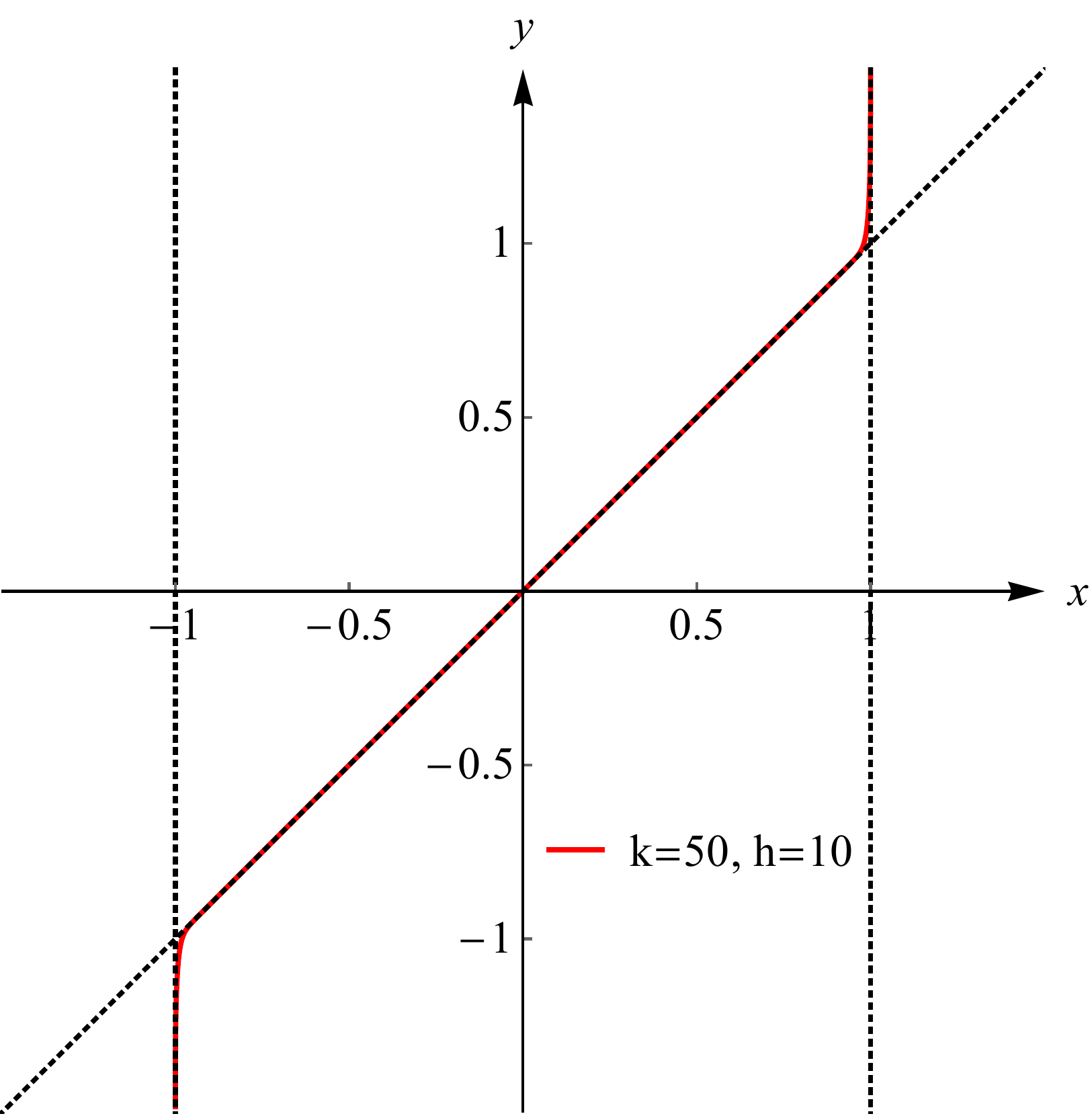}}
	\hfill
	\subfloat(b){%
		\includegraphics[width=0.17\linewidth]{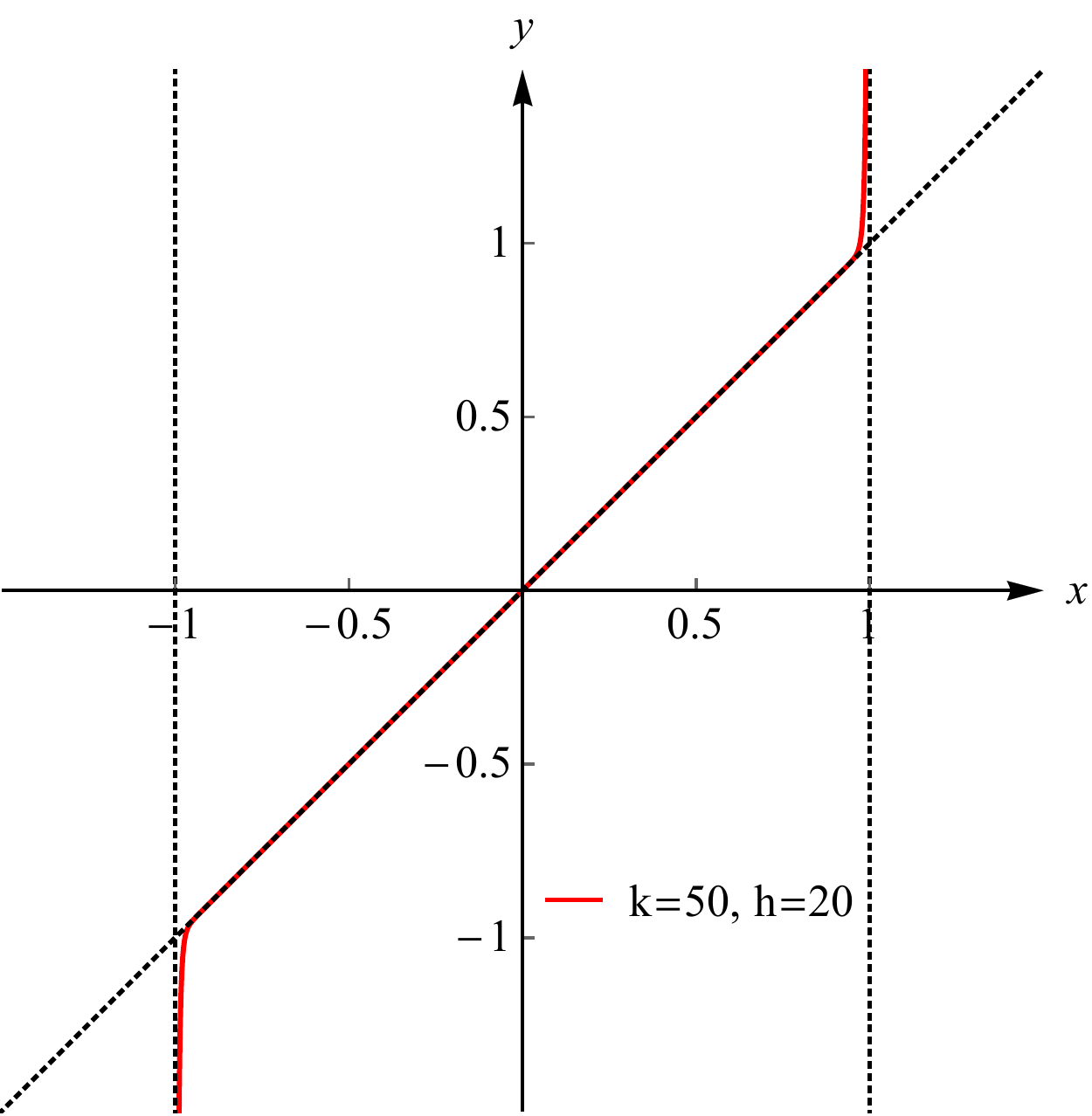}}
	\hfill
	\subfloat(c){%
		\includegraphics[width=0.17\linewidth]{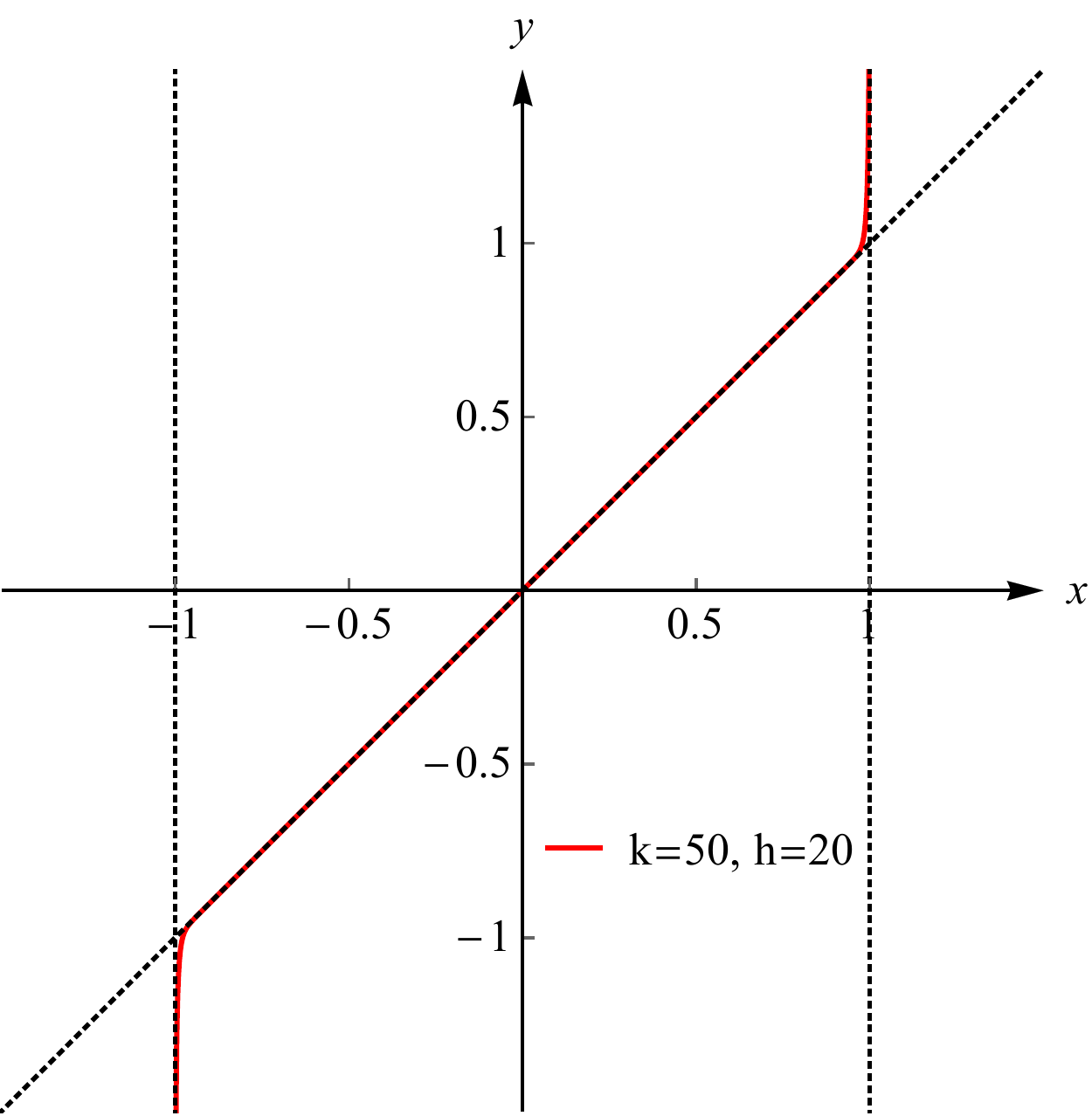}}
	\hfill
	\subfloat(d){%
		\includegraphics[width=0.17\linewidth]{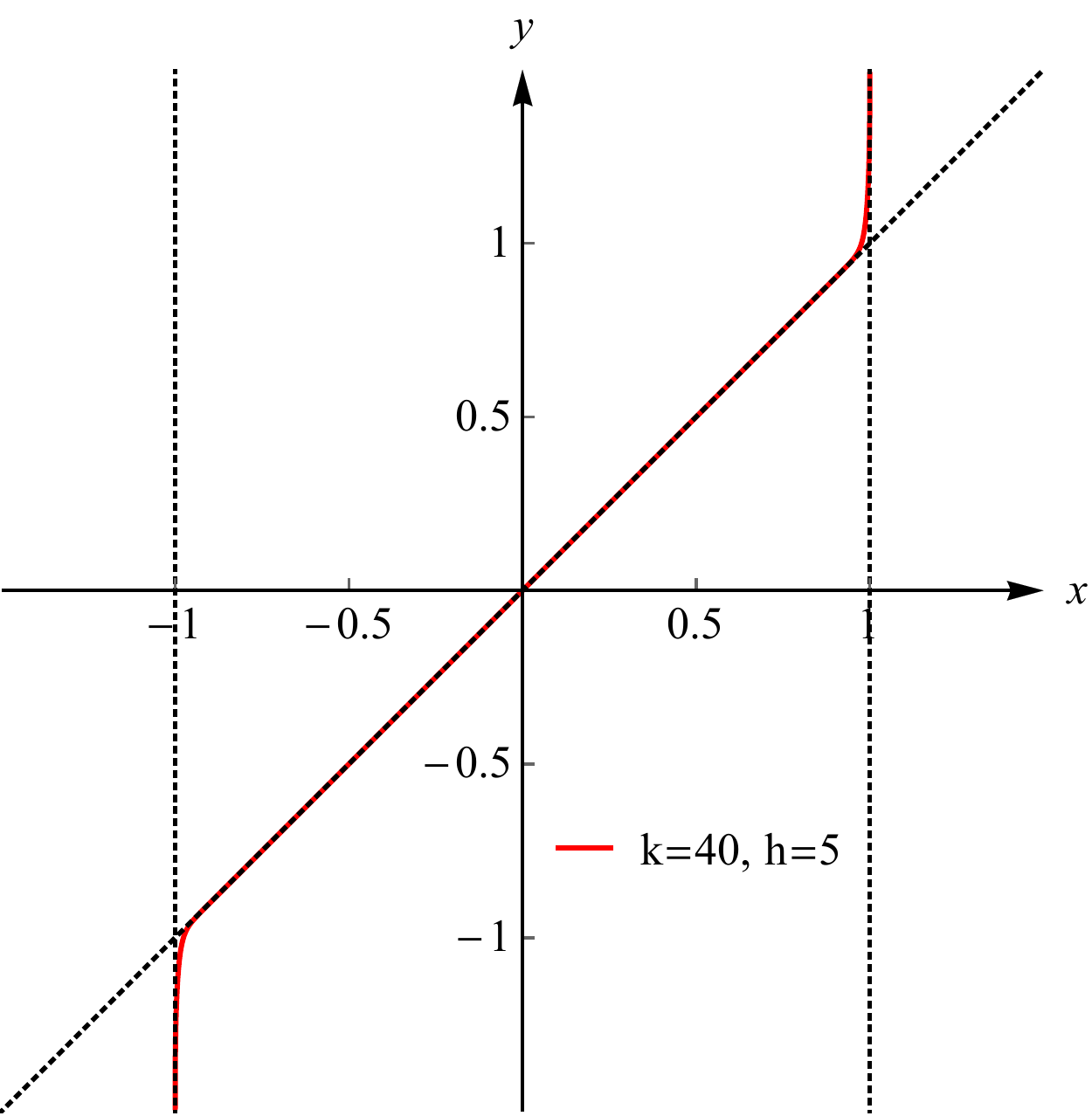}}
	\hfill
	\subfloat(e){%
		\includegraphics[width=0.17\linewidth]{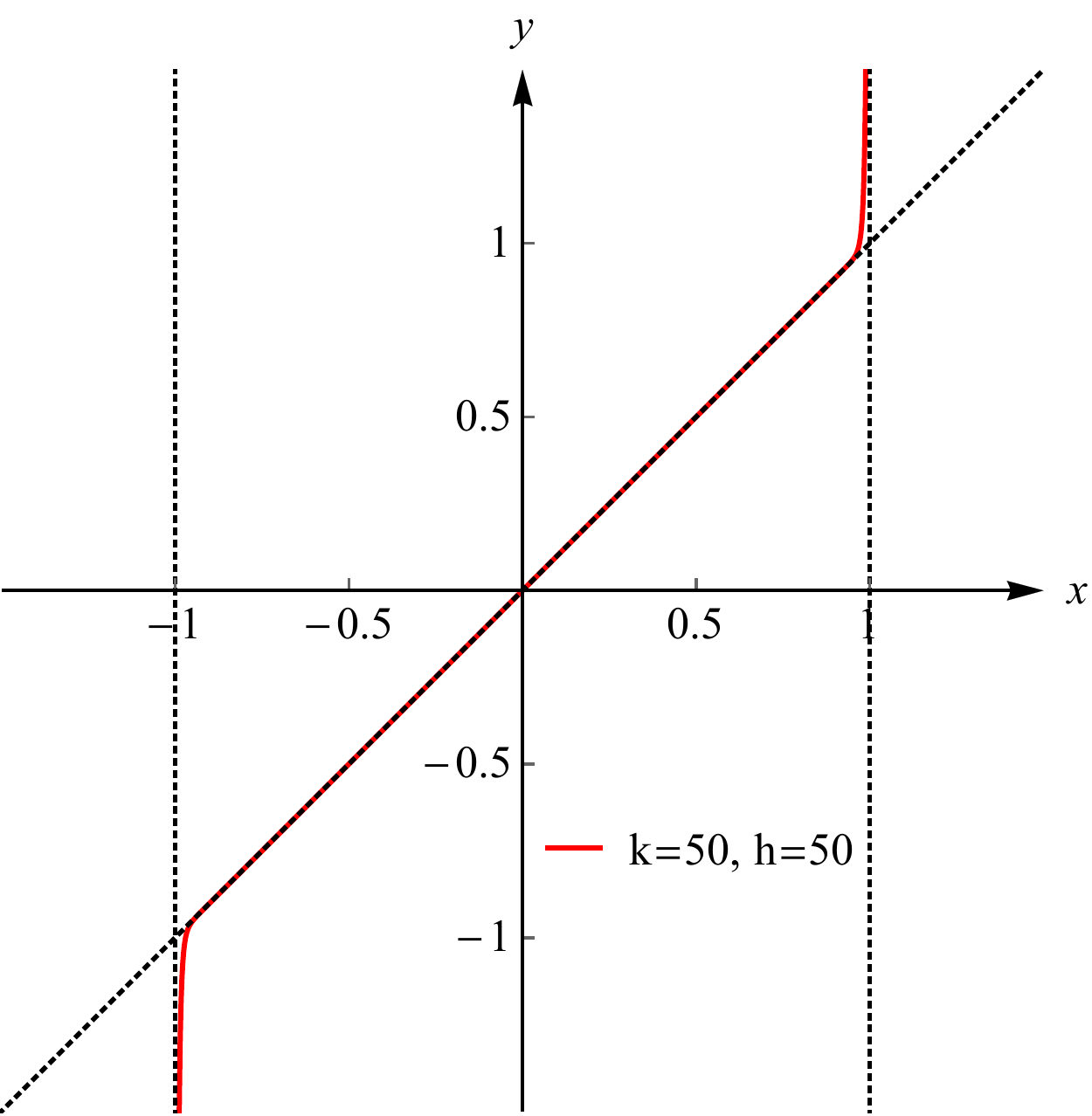}}
	\caption{The characteristics of $(x,y) \in \Fcal^{\mathrm{smooth}}_{\pi/4}$ for the $k$ and $h$ values used in simulations with the candidate curves: (a) (\ref{eq:ac1}), (b) (\ref{eq:ac2}), (c) (\ref{eq:ac3}), (d) (\ref{eq:ac4}), (e) (\ref{eq:ac5}).}
	\label{fig:fig9} 
\end{figure*}

\begin{figure*} 
	\centering
	\subfloat(a){%
		\includegraphics[width=0.17\linewidth]{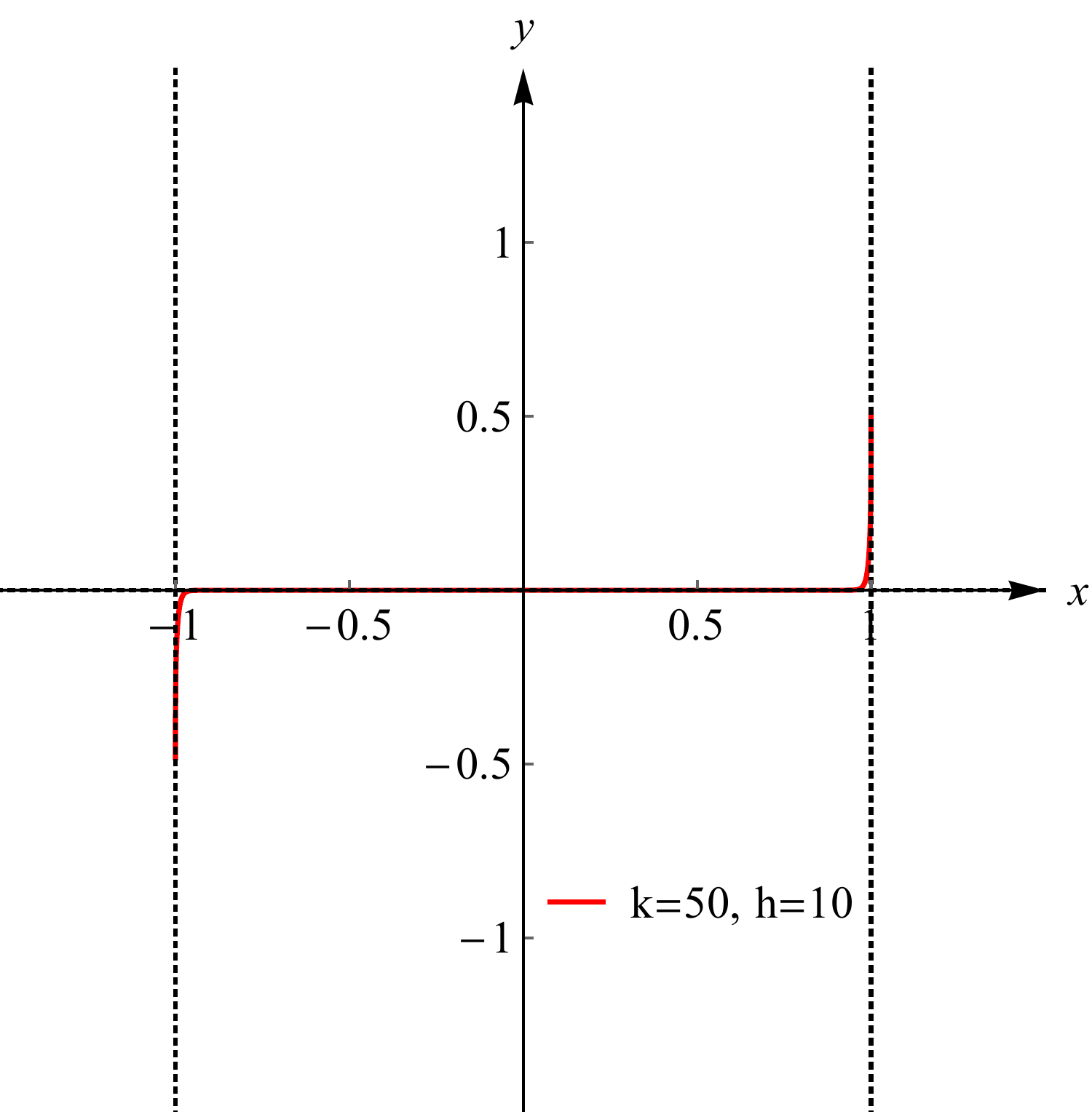}}
	\hfill
	\subfloat(b){%
		\includegraphics[width=0.17\linewidth]{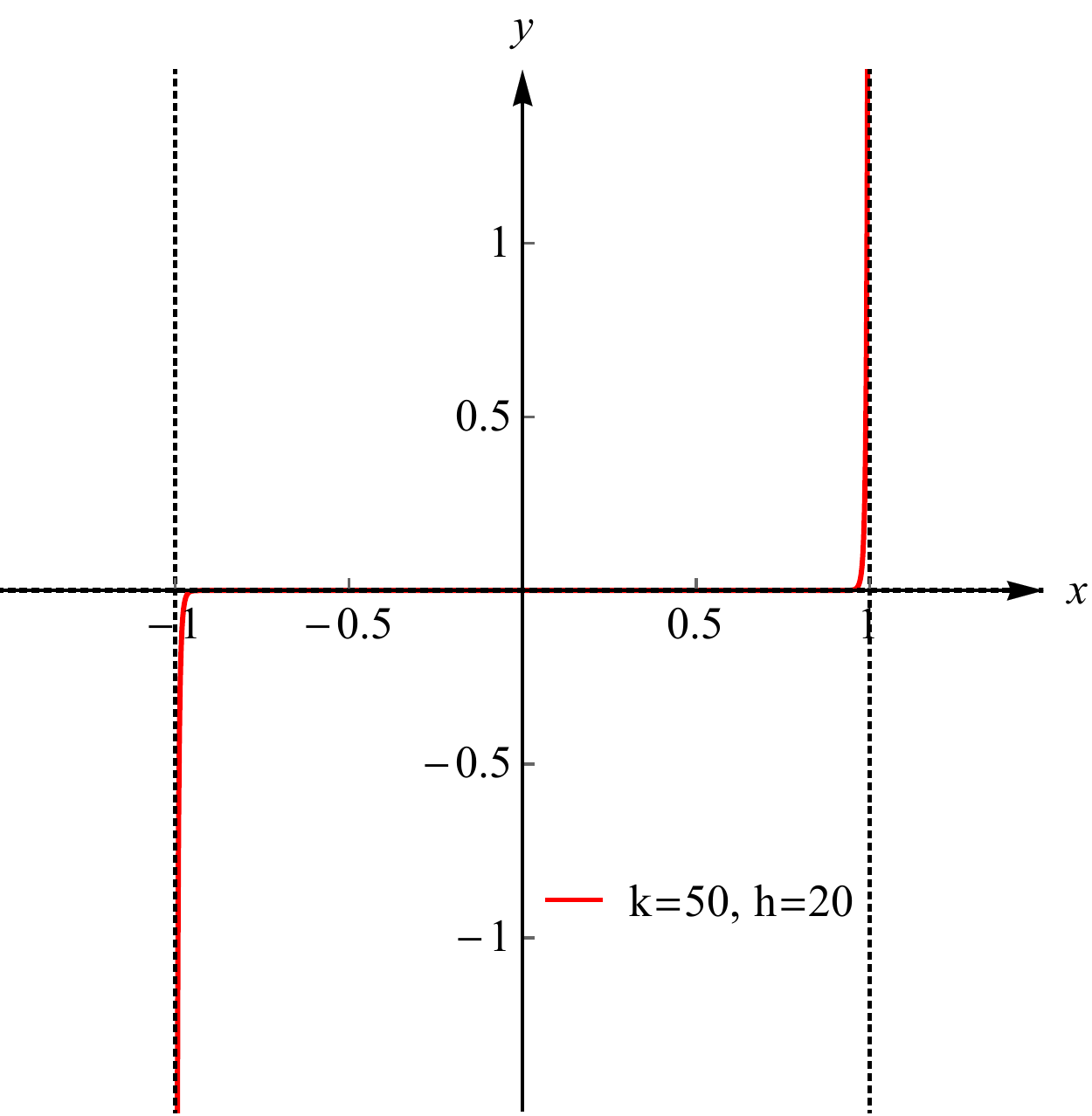}}
	\hfill
	\subfloat(c){%
		\includegraphics[width=0.17\linewidth]{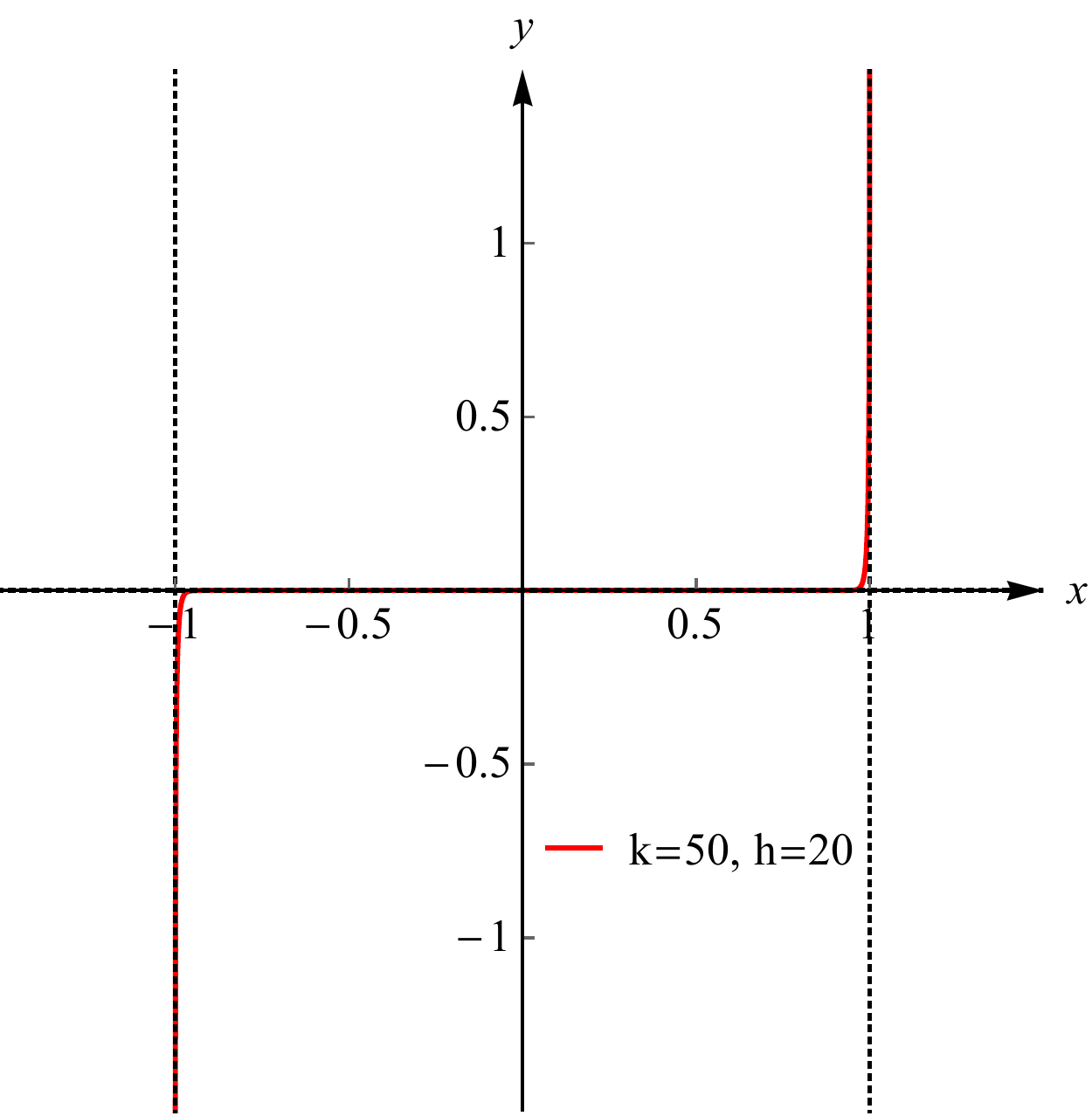}}
	\hfill
	\subfloat(d){%
		\includegraphics[width=0.17\linewidth]{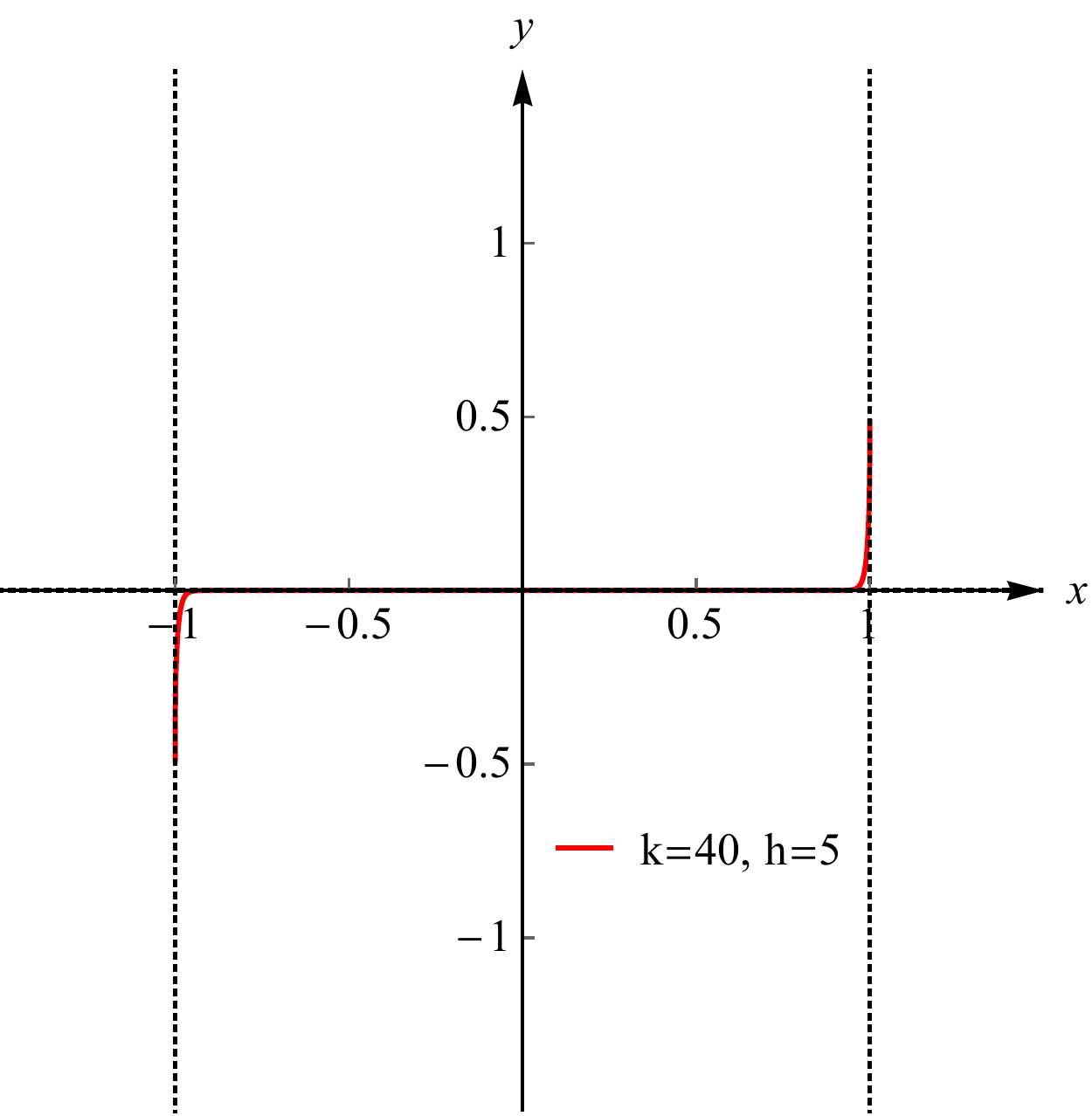}}
	\hfill
	\subfloat(e){%
		\includegraphics[width=0.17\linewidth]{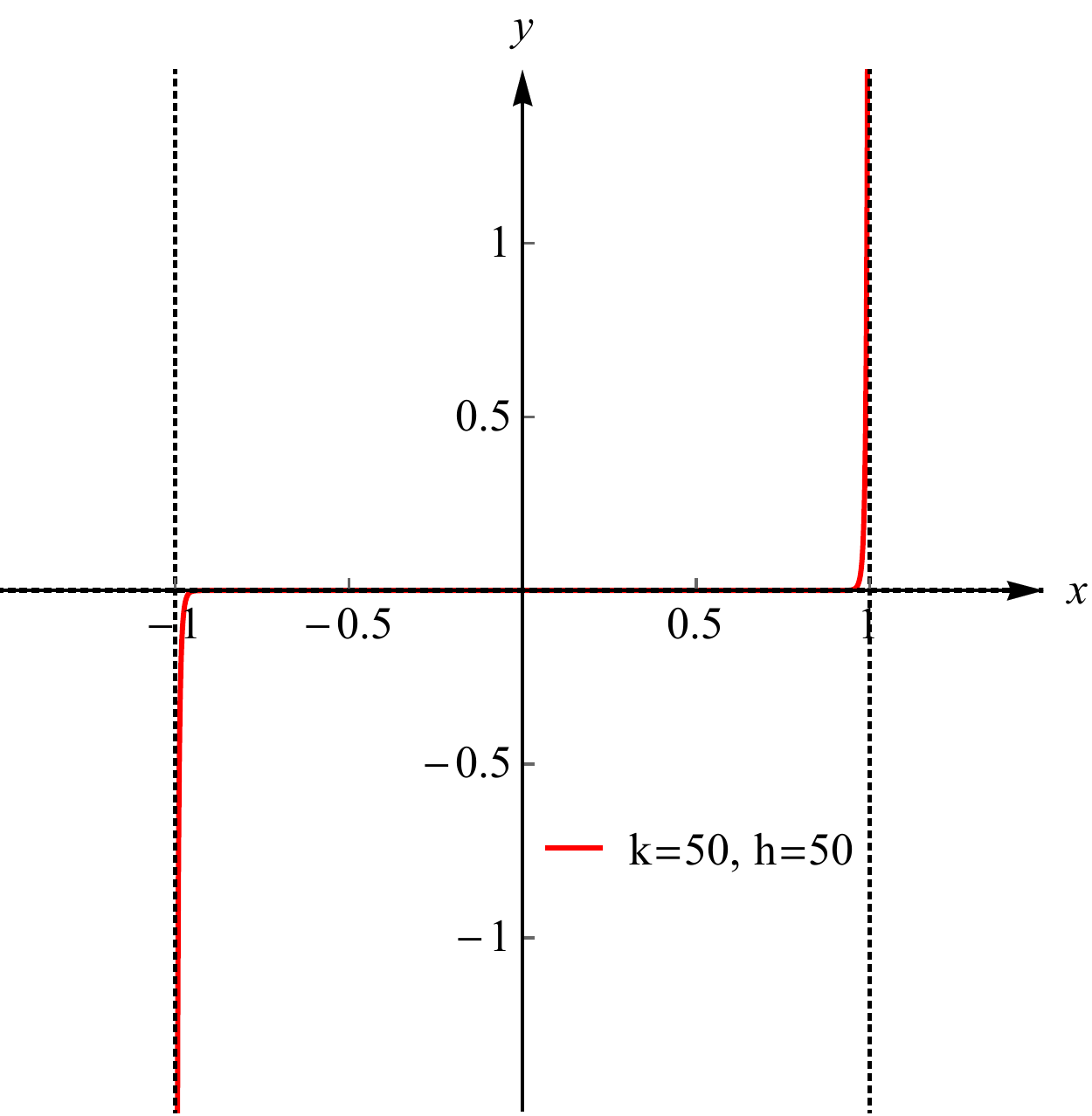}}
	\caption{The characteristics of $(x,y) \in \Fcal^{\mathrm{smooth}}_{0}$ for the $k$ and $h$ values used in simulations with the candidate curves: (a) (\ref{eq:ac1}), (b) (\ref{eq:ac2}), (c) (\ref{eq:ac3}), (d) (\ref{eq:ac4}), (e) (\ref{eq:ac5}).}
	\label{fig:fig10} 
\end{figure*}

\begin{definition}
Assume that $a:\Rbb\to(-1,1)$ is an arbitrary odd, continuously differentiable, and monotonically increasing function that satisfies
\begin{align}
\lim_{x\to1} a(x)=\infty.
\end{align}
Let $\theta\in[0,\frac{\pi}{2})$, $h>0$, and $k$ be a non-negative integer. Define the family of curves 
\begin{align} 
\!\!\!\Fcal^{\mathrm{smooth}}_{\theta,\, a,\,h,\,k}
\triangleq\big\{\big(\frac{a(y)-y}{h}y^{2k}\!+\!\tan(\theta)y,\, y\big)\!\in\!\mathbb{R}^2\;|\;y\!\in\!\Rbb\big\},\!\!\!
\end{align}
as continuously-differentiable surrogates for $\Fcal_{\theta}$.
\end{definition}
Constants $h$ and $k$ offer a trade-off between the resemblance with $\Fcal_{\theta}$ and smoothness. This is demonstrated by Figures \ref{fig:fig5}, \ref{fig:fig6}, \ref{fig:fig7} and \ref{fig:fig8}. Constants $\theta$ and $k$ denote the angle of curve and the approximation order at the origin. Figures \ref{fig:fig9} and \ref{fig:fig10} illustrate the resemblance between generator's actual contingency response and candidate curve that are obtained using (\ref{eq:14}).  

Given an appropriate sigmoid and tuning parameters, one can substitute constraints (\ref{eq:OPFLC_cons_10}) and (\ref{eq:OPFLC_cons_11}) with the followings:
\begin{subequations}
{\small
	\begin{align} 
	&\!\Big(\frac{2p^{\mathrm{gen}}_{c,g}\!-\!p^{\max}_{c,g}\!+\!p^{\min}_{c,g}}{p^{\max}_{c,g}+p^{\min}_{c,g}},\frac{2(p^{\mathrm{gen}}_{0,g}\!+\!\alpha_{c,g}\Delta_c)\!-\!p^{\max}_{c,g}\!+\!p^{\min}_{c,g}} {p^{\max}_{c,g}+p^{\min}_{c,g}}\Big)
	\!\!\in\! \Fcal^{\mathrm{smooth}}_{\!\frac{\pi}{2},a,h,k}  \\
	&\!\Big(\frac{2q^{\mathrm{gen}}_{c,g}-q^{\max}_{c,g}-q^{\min}_{c,g}}
	{q^{\max}_{c,g}-q^{\min}_{c,g}}, |v_{0,i}|-|v_{c,i}|\Big) \in \Fcal^{\mathrm{smooth}}_{0,\, a,\,h,\,k}.
	\end{align}}
\end{subequations}


The candidate curves given in (\ref{eq:14}) include inverse hyperbolic tangent (\ref{eq:ac1}), inverse arctangent (\ref{eq:ac2}), inverse algebraic (\ref{eq:ac3}), inverse error (\ref{eq:ac4}), and inverse absolute-value functions. In the following section, we examine the merits of the proposed surrogate functions on the performance of local search algorithms for SCOPF.




	\section{Case Studies}
Numerical studies on the IEEE 30-bus, 118-bus, and 300-bus systems demonstrate the efficacy of the continuously-differentiable SCOPF model. The main purpose is to minimize pre-contingency power generation cost while securing the post-contingency operation. The secure operation prevents physical and operational violations by means of optimally, allocating active power imbalances among available generators and deciding the type of bus, i.e., PV/PQ switching, respectively. To this end, we consider several numerical experiments on the IEEE benchmark under an extensive list of contingency scenarios, each representing the outage of randomly chosen components, e.g., generator or line. Herein, it is ensured that randomly-chosen lines do not lead to an islanding in the grid. These studies are examined on a PC with 16-core, Xeon processor, and 256 GB RAM using Artelys Knitro v12.2.2. The allowable feasibility violation is set to $10^{-6}$ for the obtained solution of given scenarios in Table~\ref{tab:1} and \ref{tab:300}. The $h$ and $k$ values used in the simulations are given and demonstrated by Figures \ref{fig:fig9} and \ref{fig:fig10}. For all of the case studies, we also simulated the continuous complementarity model from \cite{l11}, which failed to converge to a feasible point.
     
\subsection{IEEE 30-bus System with 12 Contingencies}
The IEEE benchmark considered has 30 buses connected with 41 transmission lines, 6 generators, and 20 loads. In this benchmark, bus 1 is assigned as the slack bus; buses 2, 13, 22, 23, 27 are PV buses, and the rest are PQ buses. Consider this IEEE 30-bus system under 12 contingency scenarios as described in Table~\ref{tab:1}. To maintain system reliability under certain contingencies, the benchmark data is changed by reducing the load demand in half. Hence, the solution to the OPF problem with no contingency results in a minimum generation cost of 2847.8. Considering contingency cases in Table~\ref{tab:1} using (\ref{eq:ac1}) increases the cost during normal conditions by $11.24\%$ to 3167.9. Tables~\ref{tab:2}-\ref{tab:4} represent active power, reactive power, and voltage variations of the power grid in response to given contingencies. The proposed method distributes network's active power imbalance complying with (\ref{eq:activep}), which is due to the outage of a network component, among available generators given in Table~\ref{tab:2}. It can be observed from Table~\ref{tab:2} that the outage of a transmission line, scenarios 7-12, does not cause a considerable power redispatch. Table~\ref{tab:3} and ~\ref{tab:4} represent the reactive power and voltage relations in the case of a network component outage. It is expected that the PV buses maintain their base case voltage levels during contingencies as much as their capacity limits permit. The highlighted values in \textbf{bold} in Table~\ref{tab:4} for the scenarios 1-6, refer to the PV/PQ switching due to a generator outage. It should be noted that the outage of a transmission line, for the scenarios 7-12, does not usually require PV/PQ switching except for the bus 1 as reported in Table~\ref{tab:4}. 

Table~\ref{tab:5} compares the performance of sigmoid functions, (\ref{eq:ac1})-(\ref{eq:ac5}), in terms of the objective value and required time to solve a given problem. It can be inferred that (\ref{eq:ac1}) and (\ref{eq:ac4}) offer the best approximations in terms of objective values while others, (\ref{eq:ac2}), (\ref{eq:ac3}), (\ref{eq:ac5}) successfully recover the feasible solution considering the generator response in the case of a randomly-outed component.

\begin{table}[h]
	\centering
	\caption{List of Contingencies for IEEE 30-bus system}
			\resizebox{0.6\columnwidth}{!}{
				
	\begin{tabular}{|l|c|c|}
		\hline
		\textbf{Contingency} & \textbf{Generator} & \textbf{Line} \\
		\textbf{\;\;\;\;Number} & \textbf{Number} & \textbf{Number}\\
		\hline
		\;\;\;\;\;\;\;\;\;1& 5 & - \\
		\hline
		\;\;\;\;\;\;\;\;\;2& 6 & - \\
		\hline
		\;\;\;\;\;\;\;\;\;3& 1 & - \\
		\hline
		\;\;\;\;\;\;\;\;\;4& 4 & - \\
		\hline
		\;\;\;\;\;\;\;\;\;5& 3 & - \\
		\hline
		\;\;\;\;\;\;\;\;\;6& 2 & - \\
		\hline
		\;\;\;\;\;\;\;\;\;7& - & 2\\ 
		\hline
		\;\;\;\;\;\;\;\;\;8& - & 25\\
		\hline
		\;\;\;\;\;\;\;\;\;9& - & 20\\ 
		\hline
		\;\;\;\;\;\;\;\;10& - & 35 \\
		\hline
		\;\;\;\;\;\;\;\;11& - & 9 \\ 
		\hline
		\;\;\;\;\;\;\;\;12& - & 1\\
		\hline
	\end{tabular}}
\label{tab:1}
\end{table}

\begin{table*}
	\caption{Active power variations in response to contingency scenarios (pu)}
	\centering
	\resizebox{2\columnwidth}{!}{
		\begin{tabular}{|l||*{13}{c|}}\hline
			\diagbox[innerwidth = 1.9cm, height = 5.8ex]{Bus Nr.}{Sc. Nr.}
			&\makebox[3em]{0}&\makebox[3em]{1}&\makebox[3em]{2}
			&\makebox[3em]{3}&\makebox[3em]{4}&\makebox[3em]{5}
			&\makebox[3em]{6}&\makebox[3em]{7}&\makebox[3em]{8}
			&\makebox[3em]{9}&\makebox[3em]{10}&\makebox[3em]{11}
			&\makebox[3em]{12}\\\hline\hline
			\;\;\;\;\;\;\;\;\;1 &0.2880  &  0.2946 &   0.2945   &      0  &  0.2894  &  0.3152  &  0.3541  &  0.2878   &  0.2875  &  0.2874  &  0.2874  &  0.2874  &  0.2875 \\\hline 
			\;\;\;\;\;\;\;\;\;2 &0.4073  &  0.4131 &   0.4130   & 0.4488   & 0.4085  &  0.4311 &        0  &  0.4072 &   0.4069  &  0.4068  &  0.4068 &   0.4068 &   0.4069 \\\hline 
 			\;\;\;\;\;\;\;\;\;22&0.1755 &   0.1788  &  0.1788  &  0.1992 &   0.1762   &      0 &   0.2086  &  0.1755  &  0.1753 &   0.1752  &  0.1752  &  0.1752  &  0.1753\\\hline 
			\;\;\;\;\;\;\;\;\;27&0.0077 &   0.0208 &   0.0206  &  0.0850   &      0  &  0.0528&   0.1152 &   0.0109 &   0.0105  &  0.0102  &  0.0103 &   0.0103  &  0.0105 \\\hline 
			\;\;\;\;\;\;\;\;\;23&0.0386 &       0  &  0.0484 &   0.1096  &  0.0408  &  0.0794  &  0.1376 &   0.0385 &  0.0381 &   0.0379  &  0.0379  &  0.0379 &   0.0381 \\\hline
			\;\;\;\;\;\;\;\;\;13&0.0375 &   0.0478   &      0  &  0.1085 &   0.0401 &   0.0783   & 0.1365  &  0.0378 &  0.0374  &  0.0371 &   0.0371  &  0.0371 &   0.0374 \\\hline
	\end{tabular}}
	\label{tab:2}
\end{table*}

\begin{table*}
	\caption{Reactive power variations in response to contingency scenarios (pu)}
	\centering
	\resizebox{2\columnwidth}{!}{
		\begin{tabular}{|l||*{13}{c|}}\hline
			\diagbox[innerwidth = 1.9cm, height = 5.8ex]{Bus Nr.}{Sc. Nr.}
			&\makebox[3em]{0}&\makebox[3em]{1}&\makebox[3em]{2}
			&\makebox[3em]{3}&\makebox[3em]{4}&\makebox[3em]{5}
			&\makebox[3em]{6}&\makebox[3em]{7}&\makebox[3em]{8}
			&\makebox[3em]{9}&\makebox[3em]{10}&\makebox[3em]{11}
			&\makebox[3em]{12}\\\hline\hline
			\;\;\;\;\;\;\;\;\;1 & -0.0888  & -0.0298  & -0.0216   &      0  & -0.0257  & -0.0245 &   0.0243 &  -0.0599 &  -0.0296 &  -0.0298 &  -0.0299  & -0.0316 &  -0.0306 \\\hline 
			\;\;\;\;\;\;\;\;\;2 & 0.1016 &   0.0500 &   0.0861  &  0.0345  &  0.0743 &   0.1011 &        0  &  0.0795 &   0.0454 &   0.0463 &   0.0459 &   0.0674 &  0.0886 \\\hline 
			\;\;\;\;\;\;\;\;\;22&0.1358 &   0.1554  &  0.1736  &  0.1425 &   0.1603  &       0  &  0.1587 &   0.1447 &   0.1164  &  0.1340  &  0.1470  &  0.1301 &  0.1337\\\hline 
			\;\;\;\;\;\;\;\;\;27&0.0748  &  0.0762  &  0.0750  &  0.0604  &       0 &   0.0905  &  0.0628  &  0.0789 &   0.0729 &   0.0732  &  0.0509 &   0.0709 &   0.0728\\\hline 
			\;\;\;\;\;\;\;\;\;23&0.0356   &      0  &  0.0817 &   0.0024  &  0.0498 &   0.0812  & -0.0074  &  0.0392 &   0.0520  &  0.0361 &   0.0465 &   0.0354 &   0.0355 \\\hline
			\;\;\;\;\;\;\;\;\;13&0.1360  &  0.1450   &      0  &  0.1375  &  0.1371  &  0.1569 &   0.1443 &   0.1428 &  0.1399  &  0.1345   & 0.1339  &  0.1333  &  0.1336 \\\hline
	\end{tabular}}
	\label{tab:3}
\end{table*}

\begin{table*}
	\caption{Voltage variations in response to contingency scenarios (pu). Values in red color indicate PV/PQ switching.}
	\centering
	\resizebox{2\columnwidth}{!}{
		\begin{tabular}{|l||*{13}{c|}}\hline
			\diagbox[innerwidth = 1.9cm, height = 5.8ex]{Bus Nr.}{Sc. Nr.}
			&\makebox[3em]{0}&\makebox[3em]{1}&\makebox[3em]{2}
			&\makebox[3em]{3}&\makebox[3em]{4}&\makebox[3em]{5}
			&\makebox[3em]{6}&\makebox[3em]{7}&\makebox[3em]{8}
			&\makebox[3em]{9}&\makebox[3em]{10}&\makebox[3em]{11}
			&\makebox[3em]{12}\\\hline\hline
			\;\;\;\;\;\;\;\;\;1 & 1.0465  &  {\textbf{1.0489}} &   {\textbf{1.0487}}  &  1.0466 &   {\textbf{1.0488}}  &  {\textbf{1.0488}} &   {\textbf{1.0477}}  &  {\textbf{1.0500}}  &  {\textbf{1.0489}}  &  {\textbf{1.0489}} &   {\textbf{1.0489}}  &  {\textbf{1.0490}} &   {\textbf{1.0490}}\\\hline 
			\;\;\;\;\;\;\;\;\;2 & 1.0470  &  1.0470  &  1.0470 &   1.0470 &   1.0470 &  1.0470  &  {\textbf{1.0409}}  &  1.0470  &  1.0470  &  1.0470 &   1.0470 &   1.0470 &   1.0470 \\\hline 
			\;\;\;\;\;\;\;\;\;22&1.0487   & 1.0487  &  1.0487  &  1.0487  &  1.0487  &  {\textbf{1.0229}}  &  1.0487  &  1.0487 &   1.0487 &   1.0487 &   1.0487 &   1.0487 &   1.0487\\\hline 
			\;\;\;\;\;\;\;\;\;27&1.0488  &  1.0488  &  1.0488  &  1.0488  &  {\textbf{1.0272}}  &  1.0489  &  1.0489  &  1.0489  &  1.0489 &   1.0489 &   1.0489 &   1.0489 &   1.0489\\\hline 
			\;\;\;\;\;\;\;\;\;23&1.0470  & {\textbf{1.0373}}  &  1.0470  &  1.0471  &   1.0471 &   1.0470  &  1.0471 &   1.0471  &  1.0471 &   1.0471 &   1.0471 &   1.0471 &    1.0471 \\\hline
			\;\;\;\;\;\;\;\;\;13& 1.0664 &   1.0665 &   {\textbf{1.0325}} &   1.0665 &   1.0665 &   1.0665 &  1.0665 &   1.0665  &  1.0665 &   1.0665 &   1.0665 &   1.0665 &   1.0665 \\\hline
	\end{tabular}}
	\label{tab:4} 
\end{table*}

\begin{table}
	\centering
	\caption{Performance Comparison for the Proposed Functions}
	\resizebox{0.8\columnwidth}{!}{
		
		\begin{tabular}{|l|c|c|}
			\hline
			\textbf{\;\;\;\;\;Continuously-differentiable Function} & \textbf{Objective} & \textbf{Required} \\
			\textbf{} & \textbf{Value} & \textbf{Time (s)}\\
			\hline
			Inverse hyperbolic tangent (\ref{eq:ac1})& 3167.9 & 6.32 \\
			\hline
			Inverse arctangent (\ref{eq:ac2})& 3168.6 & 6.25 \\
			\hline
			Inverse algebraic (\ref{eq:ac3})& 3168.1 & 6.44 \\
			\hline
			Inverse error (\ref{eq:ac4})& 3167.8 & 6.86 \\
			\hline
			Inverse absolute value (\ref{eq:ac5})& 3169.1 & 7.60 \\
			\hline
	\end{tabular}}
	\label{tab:5}
\end{table}

\subsection{IEEE 300-bus System with 10 Contingencies}
Herein, we consider the IEEE 300-bus system with 10 randomly selected contingencies as indicated in Table~\ref{tab:300}. We used the inverse hyperbolic tangent function (\ref{eq:ac1}) with $k=1$ and $h=50$. The resulting model converged within 25 minutes. The OPF solution with no contingency results in the cost of $719725.11$, and adding contingencies increases the cost by $\%0.23$ to $721396.85$.

\begin{table}
	\centering
	\caption{List of Contingencies for IEEE 300-bus system}
	\resizebox{0.6\columnwidth}{!}{
		
		\begin{tabular}{|l|c|c|}
			\hline
			\textbf{Contingency} & \textbf{Generator} & \textbf{Line} \\
			\textbf{\;\;\;\;Number} & \textbf{Number} & \textbf{Number}\\
			\hline
			\;\;\;\;\;\;\;\;\;1& 28 & - \\
			\hline
			\;\;\;\;\;\;\;\;\;2& 52 & - \\
			\hline
			\;\;\;\;\;\;\;\;\;3& 7 & - \\
			\hline
			\;\;\;\;\;\;\;\;\;4& 19 & - \\
			\hline
			\;\;\;\;\;\;\;\;\;5& 35 & - \\
			\hline
			\;\;\;\;\;\;\;\;\;6& - & 214 \\
			\hline
			\;\;\;\;\;\;\;\;\;7& - & 301\\ 
			\hline
			\;\;\;\;\;\;\;\;\;8& - & 345\\
			\hline
			\;\;\;\;\;\;\;\;\;9& - & 241\\ 
			\hline
			\;\;\;\;\;\;\;\;10& - & 94 \\
			\hline
	\end{tabular}}
	\label{tab:300}
\end{table}

\subsection{IEEE 118-bus System with 100 Contingencies}
In this case study, the IEEE 118-bus system is considered with 100 uniformly chosen contingencies. Similar to the previous case study, the inverse hyperbolic tangent function (\ref{eq:ac1}) is used with $k=1$ and $h=50$. The resulting model converged within 18 hours. The solution of the OPF problem with no contingency results in the cost of $129660.70$, and the addition of contingencies leads to a $\%0.99$ increment in cost to $130946.79$.

	\section{Conclusion}

This paper tackles the SCOPF problem that considers the piecewise-discontinuous model of generator active and reactive power contingency responses by means of several continuously-differentiable models. The proposed approach provides state and decision variables to ensure continuity of power grid operation even under contingencies. The problem is expressed as a nonlinear programming formulation and involves piecewise-smooth models due to the characteristics of the active and reactive power contingency. We replace these non-differentiable curves with several continuously-differentiable surrogates that are tractable and can be solved with various numerical solvers. The proposed models are numerically verified on several IEEE benchmarks under various contingencies.

	
	\bibliographystyle{IEEEtran}
	\bibliography{paper1}
	\balance
\end{document}